\documentclass[floatfix,aps,twocolumn,floats,pre]{revtex4-1}
\usepackage{graphics,graphicx,epsfig}
\usepackage{epsf,epstopdf,wrapfig}
\usepackage{amsmath,amssymb}
\usepackage{xcolor}
\usepackage{dsfont}

\usepackage{newfloat}
\usepackage{algorithm}
\usepackage{algpseudocode}

\begin{document}

\title{Capturing the Continuous Complexity of Behavior in {\it C. elegans}}

\author{Tosif Ahamed\textsuperscript{a},
Antonio C. Costa\textsuperscript{b},
Greg J. Stephens\textsuperscript{a,b}}

\affiliation{$^a$Biological Physics Theory Unit, OIST Graduate University, Okinawa 904-0495, Japan\\
$^b$Department of Physics and Astronomy, Vrije Universiteit Amsterdam, 1081HV  Amsterdam, The Netherlands
\\
}

\begin{abstract} 
Animal behavior is often quantified through subjective, incomplete variables that may mask essential dynamics. Here, we develop a behavioral state space in which the full instantaneous state is smoothly unfolded as a combination of short-time posture dynamics. Our technique is tailored to multivariate observations and extends previous reconstructions through the use of maximal prediction. Applied to high-resolution video recordings of the roundworm \textit{C. elegans}, we discover a low-dimensional state space dominated by three sets of cyclic trajectories corresponding to the worm's basic stereotyped motifs: forward, backward, and turning locomotion. In contrast to this broad stereotypy, we find variability in the presence of locally-unstable dynamics, and this unpredictability shows signatures of deterministic chaos: a collection of unstable periodic orbits together with a positive maximal Lyapunov exponent. The full Lyapunov spectrum is symmetric with positive, chaotic exponents driving variability balanced by negative, dissipative exponents driving stereotypy. The symmetry is indicative of damped, driven Hamiltonian dynamics underlying the worm's movement control.
\end{abstract}

\maketitle

\section*{Introduction}

Animals move in a wide variety of ways; the complex posture dynamics generating these behaviors span multiple spatiotemporal scales, and exhibit both regularity and variability \cite{Gray1953,Scafetta2009}. At large scales, behavior is structured, organized into stereotyped motifs such as walking or running, but the dynamics within each motif can be highly irregular \cite{Grobstein1994,Newell1993}. This complexity is apparent in spontaneous behaviors \cite{Cole1995,Maye2007}, but also in highly stereotyped sequences such as an ``escape response" \cite{Card2012}, which must also be unpredictable for successful avoidance from motile predators \cite{Moore2017}. Despite the importance of behavior in fields ranging from neuroscience \cite{Gomez-Marin2014,Krakauer2017}, ethology \cite{Brown2018,Berman2018}, control theory \cite{Cowan2014}, robotics and artificial intelligence \cite{Aguilar2016} to the physics of living systems \cite{Ramaswamy2017active}, the complexity of movement presents unique challenges in quantification, analysis, and understanding.

Technological advances, including recent progress in machine vision \cite{Mathis:2018us,Pereira:2019te}, now make it possible to gather high-resolution movement data, even in complex, naturalistic settings and for animals with intricate body plans \cite{Han:2018jv,Wiltschko2015,Graving2019}. But how do we map high-resolution recordings of animal behavior into a compressed set of interpretable numbers while retaining maximal information about the dynamics?  Indeed, among biological signals, behavior exhibits a remarkable divergence of descriptions, from representations based on pixels and wavelets (see e.g.~\cite{Berman:2014ba}), to postures (see e.g. \cite{stephens2008ploscomp}) to more abstract states (see e.g.~\cite{Flavell:2013bp,Johnson2019}). Certainly, a good representation should capture the difference between distinct movement patterns. An ideal representation will also allow near-future predictions and be interpretable so as to provide insight into movement control principles. Finally, we seek to reveal rather than impose on the structure of the behavioral signal, letting the representation and analysis guide important characteristics such as continuous vs.~discrete, variable vs.~stereotyped and spontaneous vs.~controlled. 

We detail the construction and application of a {\it behavioral state space} inspired by the similar approach of dynamical systems (also known as a phase space \cite{Strogatz2018,Nolte2010}, not to be confused with ``state-space models'' in statistics \cite{Durbin2012}). A point in our generally multidimensional behavioral state space represents the complete, near-instantaneous movements of an animal: posture and short-time posture changes. As time evolves, the state-space point follows a smooth trajectory, thus providing a geometrical encoding of behavior. Combining dynamical systems theory with high-resolution posture time series of the nematode {\em C. elegans}, we exploit the detailed structure of these trajectory encodings to seek a new quantitative perspective of ethological analysis \cite{Tinbergen1963}.  

\section*{State Space Reconstruction by Maximizing Predictability}
\begin{figure*}
\begin{center}

\includegraphics[width=0.9\textwidth]{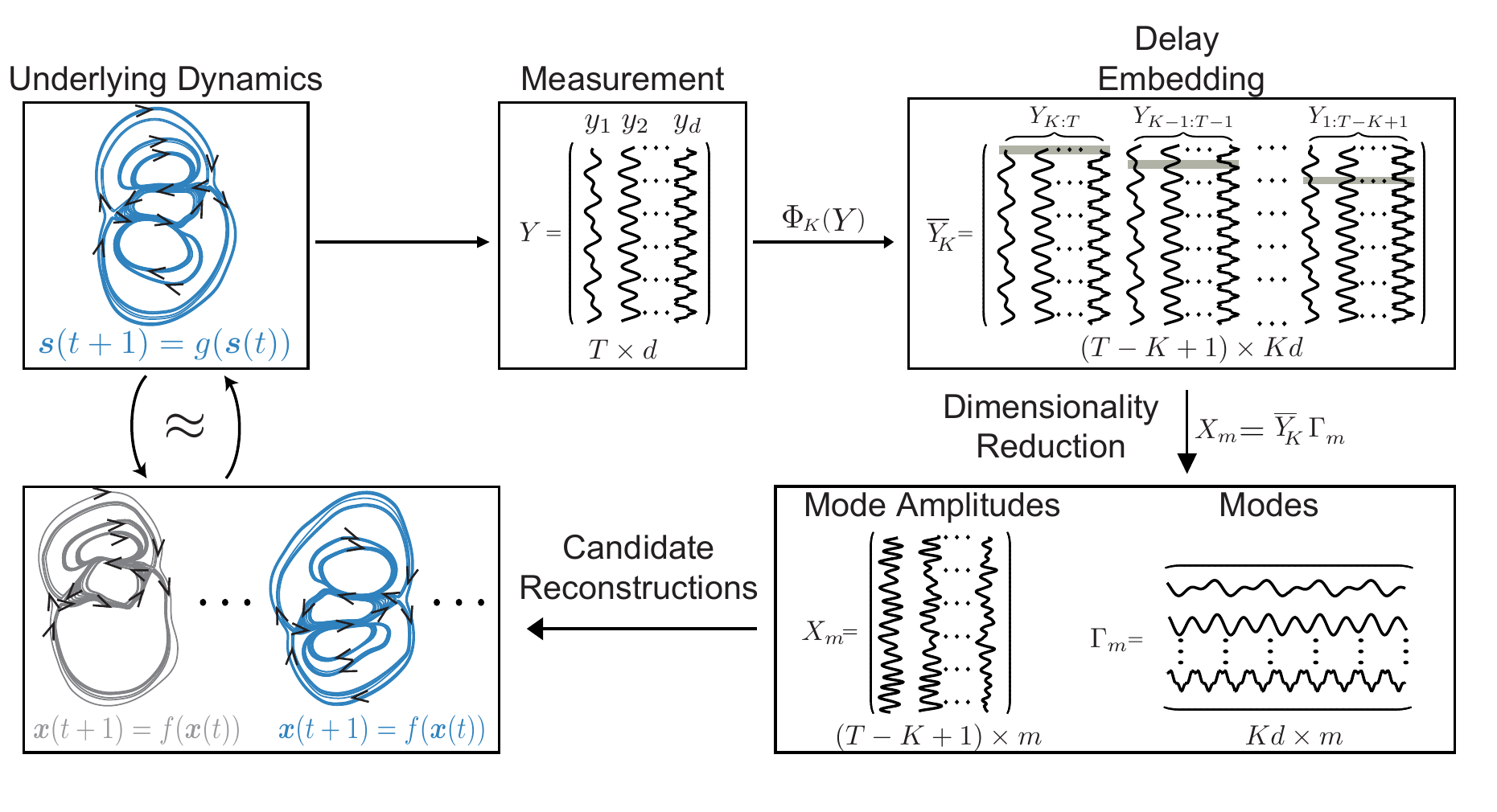}
\caption{{\bf State Space Reconstruction.}
{\bf (From Upper Left)} A $d$-dimensional time series of an underlying dynamical system is collected in the measurement matrix $Y$. The embedding operation $\Phi_{K}$ stacks delayed copies of the measurements within a short time window of length $K$ into a matrix $\overline{Y}$ of size $(T-K+1) \times Kd$. Dimensionality reduction of $\overline{Y}$ results in an $m<Kd$ dimensional state space spanned by the basis vectors $\Gamma_{m}$ called the modes. The mode coefficients $X_{m}$ form an approximation of the state space of the underlying dynamics. Each value of $K$ and $m$ results in a different state space reconstruction of the underlying dynamics and we seek embedding parameters which maximize predictability. 
}
\label{fig:fig1}
\end{center}
\end{figure*}

We consider a $d-$dimensional time series $\vec{y}(t)$ of duration $T$ collected in a $T\times d$ matrix ${Y}$, which represents noisy, incomplete measurements of an underlying dynamical system, Fig.~\ref{fig:fig1}. With a state space reconstruction, we seek a coordinate transformation $\Psi$ that maps $Y$ into a space $X$ that is topologically equivalent \cite{Palis1982} to the state space of the underlying dynamical system, a process known as time series embedding \cite{Casdagli1991,broomhead86}. Dynamical embeddings have been used to model complex phenomena such as ecological and neural dynamics \cite{Sugihara1990,Tajima_2015}, and to characterize the stability and symmetry of their reconstructed attractors \cite{Kantz2004}. Although early approaches primarily used singe-variable measurements, multivariate embeddings provide better reconstructions \cite{read1993phase} and can improve prediction \cite{Ye2016}.

In our approach, we first lift the $d$-dimensional measurements into a $Kd$ dimensional space of $K$ contiguous delays and then project to a smaller $m$-dimensional subspace. Formally we decompose the embedding $\Psi=P_{m} \circ \Phi_{K}$ into a delay map $\Phi_{K}$ in which we iteratively stack $(K-1)$-delayed copies of $Y$ into a $(T-K+1) \times Kd$ matrix $\overline{Y}_K$, followed by a dimensionality reduction transformation $P_{m}$ which projects $\overline{Y}_K$ onto an $m<Kd$ dimensional space. $P_{m}$ can in principle be any transformation and examples include numerical derivatives \cite{Packard1980}, delay coordinates \cite{Takens1981} and random projections \cite{Tajima_2015}. Here, we use singular value decomposition (SVD) \cite{broomhead86,read1993phase} followed by independent components analysis (ICA) \cite{Hyvaerinen1999}, which results in a state space with independent components spanning the dimensions of the first $m$ singular vectors. In matrix notation $X_{m}=\overline{Y}_K \Gamma_{m}$, where $\Gamma_{m}$ is the $Kd\times m$ matrix of basis vectors spanning the $m$ dimensional state space, while $X_{m}$ contains the state space trajectories. This space of transformations allows for both derivative and more general linear filters \cite{Gibson1992} and the resulting coordinates reflect the most significant linear modes of the dynamics \cite{Gibson1992,broomhead86}. 
 
\begin{figure*}
	\begin{center}
		\includegraphics[width=0.9\textwidth]{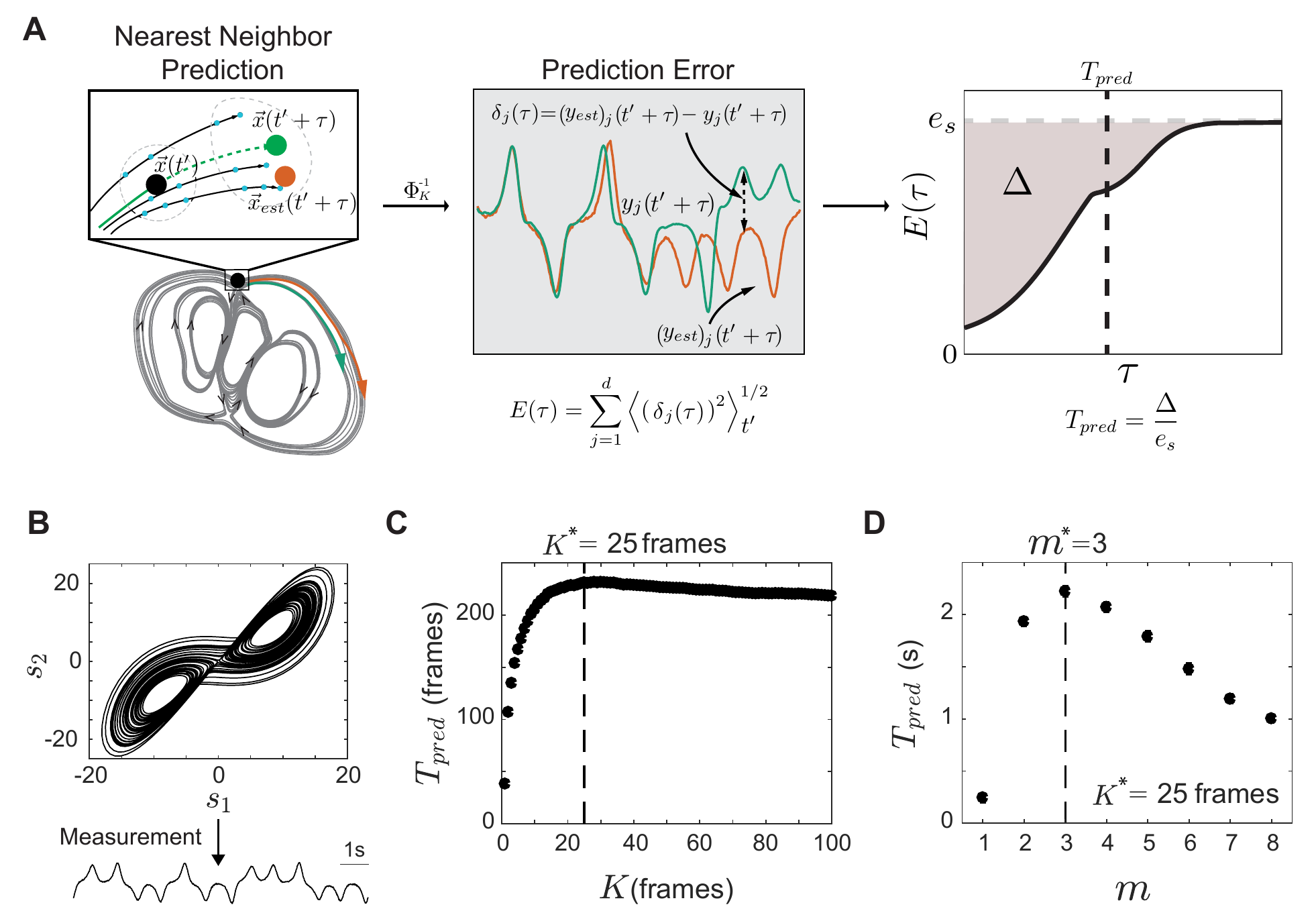}
		\caption{{\bf Reconstruction Parameters for Maximum Predictability.} \textbf{(A)} We use a nearest neighbor predictor, which averages the $\tau$-step future of nearest neighbors of a point $\textbf{x}(t)$ in the candidate reconstruction to estimate future measurements. $E(\tau)$ (Eq.~\ref{eq:error}) quantifies the error in prediction as a function of $\tau$. In general for stationary complex systems, prediction gets worse with time and $E(\tau )$ grows before saturating to a value $e_{s}$, corresponding to the size of the system in state space. We define the characteristic timescale for $E(\tau)$ to saturate to $e_s$ as a new measure of predictability, $T_{pred}$ (Eq.~\ref{eq:tpred}), geometrically given by the area between $e_{s}$ and $E( \tau )$ divided by $e_{s}$. \textbf{(B-D)} We apply our reconstruction to the time series corresponding to a noisy measurement of the first coordinate of the Lorenz system with standard chaotic dynamics. \textbf{(C)} The prediction time $T_{pred}$ varies for different $K$, with a maximum $K^{*}\approx0.25\ s$. \textbf{(D)}  $T_{pred}$ is maximal for the reconstruction defined by $K^*$ and the first three singular vectors, resulting in a 3D embedding of the chaotic attractor, which has fractal dimension $D\approx2.05$. Errorbars in (C,D) are comparable to the point size.
		}
		\label{fig:fig2}
	\end{center}
\end{figure*}

The reconstruction is parameterized by the window length $K$ and the state space dimension $m$ and we describe a new, principled procedure for determining $(K,m)$ based on optimal prediction.  Notably, embedding parameters have often been chosen heuristically (see e.g.~\cite{Gibson1992,broomhead86}).  To predict future observations we use $N_{\rm b}$ nearest neighbors in the reconstructed state space, Fig.~\ref{fig:fig2}A(left). To compute ${\vec{y}_{\,\rm est}}(t+\tau )$, the $\tau$-step prediction of $\vec{y}(t)$, we average the future of the nearest neighbors of the corresponding state space point  so that $\vec{x}_{\,\rm est}( t+\tau) =\langle \vec{x}(t+\tau) \rangle_{N_b}$ and then apply $\Phi_K^{-1}$ to pull $\vec{x}_{\,\rm est}$  back to observation space. This is known as the nearest neighbor predictor and also as Lorenz's ``method of analogs'' \cite{Lorenz1995}. The nearest neighbor predictor provides a lower bound to the predictability of a state space reconstruction as it is equivalent to a zeroth order Taylor approximation of the dynamics in a local neighborhood.  

We quantify the prediction quality after $\tau$ steps using the error 
\begin{equation}
E(\tau)=\sum ^{d}_{j=1} \langle \left( y_{j}\left( t^{\prime } +\tau \right) - (y_{\rm est})_j \left( t^{\prime } +\tau \right)\right)^{2} \rangle ^{1/2}_{t^{\prime }} 
\label{eq:error}
\end{equation} 
as shown in Fig.~\ref{fig:fig2}A(middle). Although $E(\tau)$ is {\it function} we seek a single scalar that captures overall predictability.  For a completely predictable system $E(\tau)$ is constant with a value corresponding to the noise level in the observations. On the other hand, for systems where predictions get worse over time, $E(\tau )$ grows according to a non-trivial process, possibly involving multiple timescales \cite{Lorenz1995,Aurell1997,Gao2006}, shown schematically in Fig.~\ref{fig:fig2}A(right). 

As long as the system is stationary, the error is bounded by the maximum distance within the state space, denoted $e_{s}$; $E( \tau )$ grows until it saturates to $e_{s}$ as $\tau \rightarrow \infty $, at which time the predictions are as good as choosing randomly from the sampled state space. We use the cumulative difference between the early-time and asymptotic error to define $T_{\rm pred}$ as a new measure of predictability, 
\begin{equation}
T_{pred} =\frac{1}{e_{s}}\int ^{\infty }_{0} (e_{s} -E(\tau ))\,d\tau = \frac{\Delta}{e_{s}}
\label{eq:tpred}
\end{equation}
where $\Delta$ is the area between the curve $E(\tau )$ and the asymptote $e_{s}$. A state space reconstruction with a large value of $T_{\rm pred}$ is good in the sense that it allows us to predict future observations for as long as possible. Although several previous studies about state space reconstruction are based on prediction as a guiding principle, they have used the predictive error in a more {\it ad hoc} manner, either by setting $\tau$ to a specific value \cite{Sugihara1990,Kennel1992,Cao1998,Casdagli1991}, or by integrating $E(\tau)$ to a chosen time $\tau _{0}$ \cite{Judd1998}. \color{black}

The average prediction error for an arbitrary time $\tau ^{\prime }$ is $\langle E(\tau') \rangle  =\frac{1}{\tau'}\int ^{\tau ^{\prime }}_{0} E(\tau )\,d\tau $. At large enough $\tau ^{\prime }$, the error $E(\tau )$ approaches $e_{s}$ and we can write  $\langle E(\tau') \rangle=e_{s} -\Delta/\tau'$. Thus, the average prediction error is reduced from its asymptotic limit by an amount given by $\Delta/\tau'$.   $T_{pred}$ is also the characteristic timescale for a ball of points to randomize according to the state space density. 
 
We demonstrate our embedding approach on a noisy measurement of a single coordinate of the Lorenz system (Methods: Lorenz System) and display the results in Fig.~\ref{fig:fig2}B-E. We find that $T_{\rm pred}$ increases with $K$ for $K<25$ frames after which it decreases gradually and we choose $K^{*}=25$. We project $\overline{Y}_{K^{*}}$ on the first $m$ singular vectors and find that $T_{pred}$ decreases after $m^{*}=3$. 

\section*{The Low-Dimensional State Space of {\em C. elegans} Locomotion}
\begin{figure*}
\begin{center}
\includegraphics[width=0.8\textwidth]{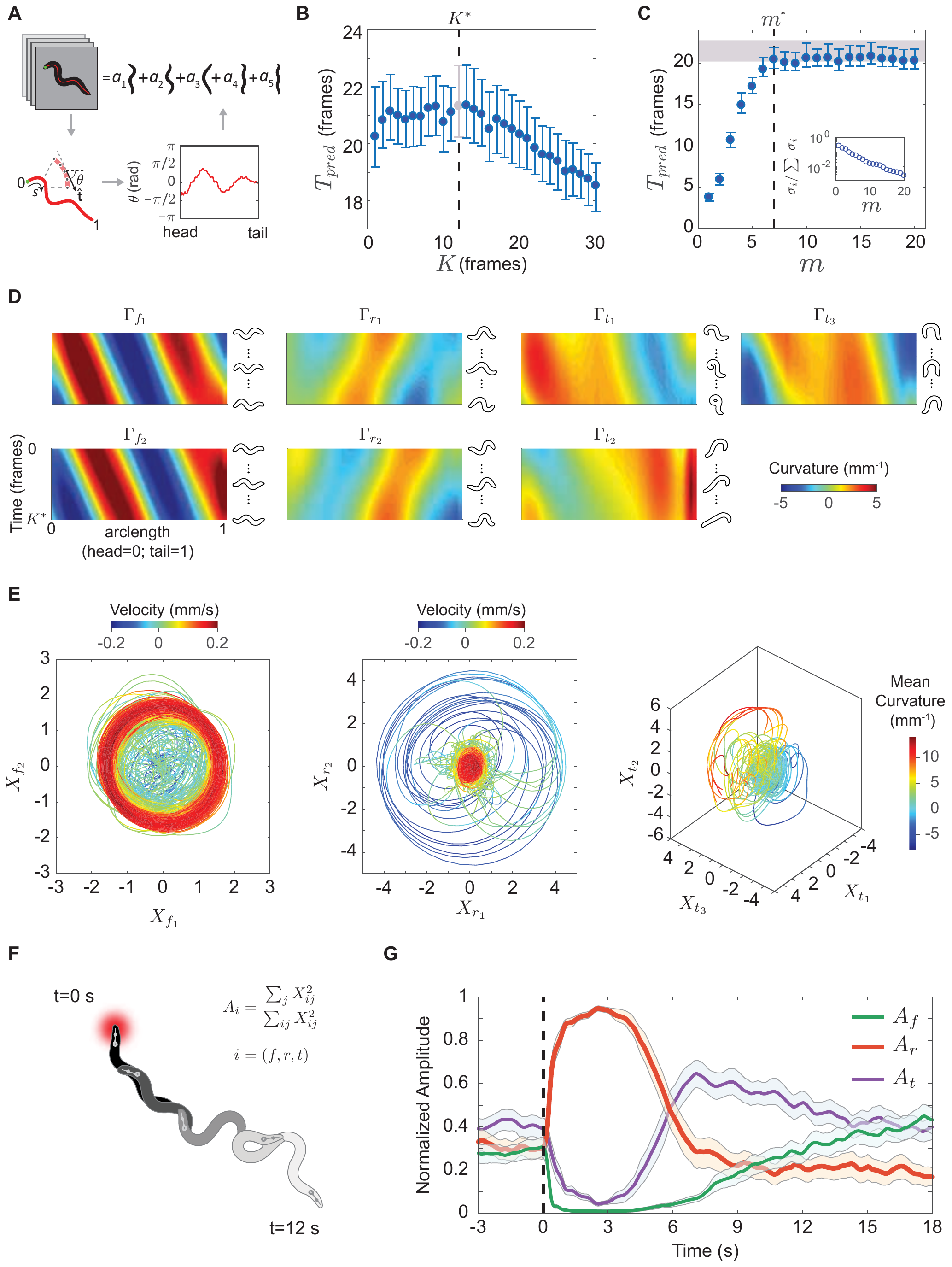}
\caption{{\bf The Low-Dimensional State Space of {\em C.elegans} Locomotion.} \textbf{(A)} We parameterize worm posture by eigenworm projections \textbf{(B-E)} The embedding process for the foraging behavior of an example worm. \textbf{(B)} $T_{pred}$ as a function of $K$. We set $K^{*}=12$, $\sim1/2$ cycle of the worm's body wave. \textbf{(C)} $T_{pred}$ as a function of $m$.  Embeddings with dimensions beyond $m^{*}=7$ carry little predictive structure. The gray bar denotes $T_{pred}(K=K^{*})$. \textbf{(C, inset)} The normalized singular value spectrum does not show an obvious cut-off. \textbf{(D)} We decompose the 7D embedding into linear combinations of independent posture sequences of length $K^{*}$, which we denote behavioral modes.  We show the modes as curvature kymographs and note that there are two approximately conjugate pairs and a third group with three modes. $\Gamma_{f1}$ and $\Gamma_{f2}$ correspond to the body waves of forward locomotion, $\Gamma_{r1}$ and $\Gamma_{r2}$ to backward locomotion, and $\Gamma_{t1}$, $\Gamma_{t2}$ and $\Gamma_{t3}$ to deep body bends. \textbf{(E)} Trajectories visualized as projections onto the above mode combinations. In the $(X_{f1},X_{f2})$, and $(X_{r1},X_{r2})$ planes, the trajectories are colored by centroid velocity, negative (blue) for backward locomotion, and positive (red) for forward. Trajectories in the $(X_{t1},X_{t2},X_{t3})$ space are colored by the mean body curvature (blue for dorsal, red for ventral). \textbf{(F-G)} We apply our embedding to the behavior of worms escaping from a heat impulse to the head. \textbf{(F)} Schematic of the response. \textbf{(G)} We project the escape dynamics using the ensemble foraging modes (Fig.~\ref{Fig:supp_ensembl_modes}) and visualize the dynamics through normalized amplitudes of the forward, reversal and turning projections. Reversal dynamics (red) initiate just after the impulse (dotted line) while turning dynamics (purple) are suppressed. The reversal ends with the initiation of an $\Omega$-turn.
}
\label{fig:fig3}
\end{center}
\end{figure*}

We leverage our state space reconstruction to elucidate the behavior of the nematode \emph{C. elegans} freely-foraging on a flat agar plate  \cite{stephens2011pnas, Broekmans2016}. In 2D, worms move by making dorsoventral sinusoidal bends along their body \cite{Croll1975, Croll1975a}, which can be captured through high resolution tracking microscopy to give a multidimensional time-series of posture changes \cite{Likitlersuang2012}. Despite the variety of visible postures, most of the shape variation is captured by a linear combination of a small number of primitive shape dimensions (eigenworms) \cite{stephens2008ploscomp,Broekmans2016}, Fig.~\ref{fig:fig3}A.

Projections along the eigenworm dimensions describe the worm's instantaneous shape and are not a direct indication of behavior, which arises from posture changes. Dynamical representations based on derivatives \cite{stephens2008ploscomp,stephens2011pnas,Costa2019}, and on sequences of postures \cite{Schwarz2015,Gomez-Marin2014,Liu2018} have been used to quantitatively explore the worm's behavior. Importantly, the low-dimensionality of the worm's shape space doesn't imply simplicity and low-dimensionality of the behavioral dynamics, and there are several signs of complexity in \emph{C. elegans} behavior, such as heavy-tailed distributions \cite{Schwarz2015}, hierarchical structure in posture sequences \cite{Gomez-Marin2016,Gupta2019}, indications of dynamical criticality in local linear approximation of the dynamics \cite{Costa2019}, as well as simultaneous presence of stereotypy and variability in posture sequences \cite{Schwarz2015,stephens2008ploscomp,stephens2011pnas}.  

To reconstruct the state space of the worm's posture dynamics, we start with a $T\times 5$ measurement matrix $Y$ consisting of 5 eigenworm coefficients for a recording of duration $T=33600 \, {\rm frames}$ (sampled at 16 Hz), Fig.~\ref{fig:fig3}A. We stack $(K-1)$ time-shifted copies of $Y$ to give the $(T-K+1) \times 5K$ state matrix $\overline{Y}_{K}$. To estimate the optimal window size, we compute $T_{\rm pred}$ for each choice of $K$, as shown in Fig.~\ref{fig:fig3}B for a single representative worm, and choose $K^{*}=12$.  Within this window, we find that predictability saturates with $m=7$ singular vectors, Fig.~\ref{fig:fig3}C. Analysis of each worm in the foraging dataset reveals a similar simplicity, Fig.~\ref{Fig:sup_wwise_tpred}. Despite it's observed complexity, worm behavior is characterized by a low-dimensional state space. 

We increase the interpretability of the worm's state space reconstruction through a final transformation to independent components.  We use the FastICA algorithm \cite{Hyvaerinen1999} on the $m=7$ projections of the delay matrix $\overline{Y}_{K^{*}}$ to obtain independent coordinate directions and we denote these coordinates behavioral modes $\vec{\Gamma}$. We show the seven behavioral modes in Fig.~\ref{fig:fig3}D as curvature kymographs and note that they come in three groups, broadly corresponding to the three coarse categories of worm movement: forward, backward and turning locomotion. Specifically, $\Gamma_{f1}$ and $\Gamma_{f2}$ modes correspond to the ventrally and dorsally initiated anterior-posterior body waves that worms make during forward locomotion. The reversal modes $\Gamma_{r1}$ and $\Gamma_{r2}$ capture the posterior-anterior body waves worms make during backward locomotion. Finally, $\{\Gamma_{t1},\Gamma_{t2}, \Gamma_{t3}\}$ correspond to longer-ranged body bends. Large projections along $\Gamma_{t1}$ and $\Gamma_{t2}$ correspond to bends directed towards the ventral or dorsal direction respectively during a delta-turn like bend \cite{Broekmans2016}, while $\Gamma_{t3}$ corresponds to an Omega-turn like bend. In this representation, worm locomotion is approximated by linearly combining these modes with time-varying amplitudes. We find similar modes for different choices of $m^*$ (Fig.~\ref{Fig:supp_w7_modes}) and also for an ensemble embedding constructed by concatenating all $N=12$ foraging organisms, Figs.~\ref{Fig:supp_ensembl_modes}-\ref{Fig:supp_ensembl_m68}. We note that the behavioral modes emerge in an unsupervised manner, with no prior information of the worm's movement. 

The topology and geometry of trajectories in the behavioral state space contain important qualitative and quantitative information about worm behavior. A 10-min trajectory is visualized in Fig.~\ref{fig:fig3}E as projections onto the three mode combinations described above. In the $(X_{f1},X_{f2})$, and $(X_{r1},X_{r2})$ planes, trajectories are colored by the centroid velocity of the worm, negative for backward locomotion and positive for forward locomotion. Trajectories in the $(X_{t1},X_{t2},X_{t3})$ space are colored by the mean body curvature. Large excitations in each of the three projections correspond to forward, backward and turning locomotion respectively. Specifically, trajectories in the $(X_{f1},X_{f2})$ plane form a prominent circular band indicating nearly constant amplitude body waves during forward locomotion. Reversals emerge as trajectories spiraling from the center to a maximum radius in $(X_{r1},X_{r2})$ plane, and then spiraling in as a reversal ends. Finally, deep body bends are represented as large transient orbits, with ventral turns and dorsal turns on opposite sides. Wild type worms have a ventral bias in their deep body bends, which is visible in the state space as a greater density of orbits on one side of 3D projection.

The state space also captures relationships between different body wave patterns. For example, we find that most reversals transition to forward by way of a deep ventral bend (Fig.~\ref{Fig:supp_3dproj}), an observation that was previously reported in the context of the escape response and pirouette reorientation sequence \cite{Gray2005,Donnelly_2013}. To quantify the relative activity of each set of body waves and the phase relationships between them, we define normalized mode amplitudes,  $A_i=\frac{\vec{X}_i\cdot \vec{X}_i}{\vec{X}\cdot \vec{X}}$,\, where $i \in \{f,r,t\}$. The $A_i$ range from $0$ to $1$ and measure the relative activity of different body wave patterns. We use these amplitudes to examine the behavior of $N=92$ on-food worms where a brief laser impulse is applied to the head, resulting in a localized thermal stimulus provoking an escape response \cite{Mohammadi2013,Broekmans2016}, shown schematically in Fig.~\ref{fig:fig3}F. We project the posture dynamics of each stimulated worm onto the ensemble modes (Fig.~\ref{Fig:supp_ensembl_modes}) and show the normalized mode amplitudes averaged across all worms, Fig.~\ref{fig:fig3}G. The amplitudes capture the timescales and phase relationships between different body wave patterns during an escape sequence. In particular, the turning modes are strongly suppressed after the initiation of the reversal, increasing gradually as the reversal ends and worms transitions into a turn. The turning amplitude then decreases, while forward amplitude increases as worms resume forward movement in the opposite direction.

\section*{Unstable Periodic Orbits and Deterministic Behavioral Variability}

The state space of worm locomotion is organized such that neighboring points correspond to similar behavioral sequences of length $K$. However, these neighboring sequences diverge with time, resulting in longer-time unpredictability, shown as an example in Fig.~\ref{fig:fig4}A.  To understand this variability we note the strong cyclic appearance of trajectories within the projections, Fig.~\ref{fig:fig3}E, suggesting that cycles play an important role. We search for periodic orbits by identifying the first recurrence times in a neighborhood \cite{Lathrop1989,Pawelzik1991,Badii1994}. Briefly, given a point
$\vec{x}(i)$ in state space, we find the smallest $k>i$ such that $\vec{x}(k)$ is in the neighborhood of $\vec{x}(i)$. The sequence $[\vec{x}( i) ,\vec{x}(i+1) ,\cdots ,\vec{x}(k)]$ is then detected as a periodic orbit of period $p=k-i$ (Methods: Periodic Orbits). Across all foraging worms, the distribution of the number of periodic orbits exhibits peaks at approximately integer multiples of a minimum period $p_{min}$ corresponding to the frequency of each worm's body wave during forward locomotion, Fig.~\ref{fig:fig4}B\,(inset). We quantify the stability of each periodic trajectory by computing its maximal Floquet exponent (Methods: Floquet Exponents). The distribution of Floquet exponents is largely positive, indicating that the worm's periodic orbits are mostly unstable, Fig.~\ref{fig:fig4}B. The unstable periodic orbits (UPOs) of worm behavior provide a longer timescale description of the movement and also a quantitative characterization of the trajectory divergence in Fig.~\ref{fig:fig4}A. We estimate the maximal Lyapunov exponent $\lambda_{max}$ by a weighted average of the Floquet exponents of periodic orbits of increasing length, weighted by $e^{-\mu_{1} p}$, where $\mu_{1}$ is the maximal Floquet exponent of the orbit, and $p$ is its period \cite{Cvitanovic1988a}. Including orbits of duration up to $p=8$, Fig.~\ref{fig:fig4}C\,(blue), provides an approximation of $\lambda_{max}$, which  agrees with direct trajectory divergence estimates averaged across all worms (gray bar, see also Fig.~\ref{Fig:supp_divplot} and Methods: Maximal  Lyapunov  Exponent). The average across random segments of the same length converges more slowly, Fig.~\ref{fig:fig4}C\,(red). 

The detected periodic orbits are interpretable in terms of commonly observed \textit{C. elegans} behaviors. Orbits with the minimum period $p_{min}$ correspond to forward and backward crawling including orbits with a dorsal or ventral bias, Fig.~\ref{Fig:supp_fwd_rev_upo}(B-C). More surprisingly, longer periodic orbits are composites, corresponding to longer time reorientation behaviors of the worm's navigation and escape strategies \cite{Pierce-Shimomura1999,Gray2005,Donnelly_2013}. In Fig.~\ref{fig:fig4}D\,(blue) we show state space trajectories of one such period-4 orbit. This orbit is composed of a reversal followed by a deep body bend, and subsequent forward movement; a posture sequence previously reported in pirouette reorientation and escape behaviors \cite{Gray2005,Donnelly_2013}. Though this periodic orbit is several body waves long, it is repeated almost exactly at different times during the recording. We show one such recurrence Fig.~\ref{fig:fig4}D\,(orange), along with the corresponding posture sequences. The presence of such UPOs suggests an intriguing view of the worm's foraging dynamics as following a complex landscape composed of unstable orbits, a picture that is rigorously correct  for chaotic systems  \cite{Banks1992,Devaney2008}. Periodic orbits have also been investigated in a number of biological systems, including neuronal activities \cite{So1998}, human electroencephalograms \cite{So1998}, crayfish photoreceptors \cite{Pierson1995,Pei1996}, as well as cardiac arrhythmias and seizures \cite{Garfinkel1992,Schiff1994,Mishra2015}.

\section*{Symmetric Lyapunov Spectrum and Damped-Driven Hamiltonian Dynamics}

While the behavior of {\em C. elegans} is simpler than most animals, the quantitative dynamics of worm posture defy a straightforward interpretation or even, as yet, a model (see e.g.~\cite{Gjorgjieva2014,Cohen2014} for reviews). There is rough stereotypy in the orbits corresponding to each behavior, but also large cycle-to-cycle variation. Such variability is linked to a positive maximal Lyapunov exponent and unstable periodic orbits, Fig.~\ref{fig:fig4}(B,C), Fig.~\ref{Fig:supp_divplot}, so that even within a ``single'' behavior such as forward crawling, each cycle is deterministically different.  To more fully illuminate this variability, we examine the dynamics along all dimensions within the state space.

In an $m$-dimensional state space, local neighborhoods are sheared by the flow and are simultaneously stretched and squeezed along different directions, dynamics which are invariantly characterized the Lyapunov exponents, $\lambda_{i=1\dots m}$.  Formally, such stretching and squeezing is described by the Jacobian  $\mathbf{J}_{\vec{x}(t)}$, which maps an $m$-dimensional spherical neighborhood to an $m$-dimensional ellipsoid. The spectrum of Lyapunov exponents is given by the infinite time average of the logarithms of the principle axes of the ellipsoid, as illustrated in Fig.~\ref{fig:fig5}A. Positive Lyapunov exponents reflect directions along which trajectory bundles expand, while negative exponents reflect shrinking directions.

The Lyapunov exponents reveal important information about the dynamics of a system (see e.g.~\cite{Pikovsky2016}). The sum of the exponents is the average dissipation rate: zero for conservative systems and negative for those with dissipation. The sum of the positive exponents bounds the metric or Kolmogorov-Sinai (KS) entropy rate \cite{Pesin2008,Young1982}, providing a principled measure of the unpredictability. In addition, the spectrum of Lyapunov exponents can reveal underlying symmetries and conservation laws. For example, continuous dynamical systems exhibit at least one zero exponent corresponding to time-translation invariance along the direction of the flow.

We compute the Lyapunov spectrum for the state space of {\em C. elegans} (Methods: Lyapunov Spectrum and Jacobian Estimation), and show bootstrapped density estimates of the $m=7$ exponents across different worms, Fig.~\ref{fig:fig5}B. We find two positive exponents, $\lambda_1 = 0.66 \, (0.62,0.69) \, s^{-1 }$, $\lambda_2 = 0.29 \, (0.26,0.32) \, s^{-1}$, and a third, near-zero exponent $\lambda_3 = 0.056 \, (-0.02,0.11) \, s^{-1 }$. The KS entropy rate is thus bounded by the sum of positive exponents as $h_{KS} \leq 1 \, (0.93,1.09) \, {\rm nats/s}$ (note that we have restored the units of nats for ease of comparison with other entropy measures). The sum of all of the Lyapunov exponents is negative, indicating that the system is dissipative with a dissipation rate of, $\sum_{i}\lambda_{i}=-0.94 \ (-1.15,-0.78) \,  s^{-1}$. Although trajectory bundles expand locally, dissipation causes them to contract as a whole and relax to an attracting manifold. We estimate the dimension of the attractor as the Kaplan-Yorke dimension $D_{KY}=5.93 \,(5.75,6.08)$ \cite{Frederickson1983}. The combination of local expansion generating variability and local contraction generating stereotypy is an essential complexity of the worm's posture dynamics. 

The Lyapunov spectrum also exhibits a striking symmetry; exponents come in conjugate pairs that sum to the same number $\alpha=-0.27 \,(-0.3,-0.24) \, s^{-1}$, Fig.~\ref{fig:fig5}B\, (inset). The entire spectrum is thus symmetric about $\displaystyle{\frac{\alpha}{2}}$ (dotted line). The symmetry is also present in 6- and 8-dimensional embeddings, Fig.~\ref{Fig:S5}. Symmetric Lyapunov spectra have been previously observed in at least two kinds of damped-driven Hamiltonian systems: coupled oscillators with viscous damping where $\alpha$ is the dissipation per degree of freedom \cite{Dressler1988}, and thermostatted molecular dynamic simulations where $\alpha$ is a feedback friction force that acts to maintain a dynamic equilibrium by either keeping the the kinetic energy of the particles constant \cite{Gupalo1994,Dettmann1996,Ruelle1999,Bright2005}.  Interestingly, in a biomechanical model of larval Drosophila locomotion, damped-driven Hamiltonian chaotic dynamics were sufficient to generate realistic forward and backward crawling, as well as more complex reorientation behaviors  \cite{Loveless2019}.

\section*{Discussion}
We use sequences of multidimensional data to reconstruct a maximally predictive state space (Fig.~\ref{fig:fig1}, Fig.~\ref{fig:fig2}). Conceptually, our approach is a timescale separation; short-time sequences define the reconstructed state variables while longer-time dynamics are encoded as state space trajectories. Our reconstruction {\em explicitly} seeks the full state information available in short-time dynamics, analogous to discovering the additional variable of velocity from the displacement time series of a simple oscillator.  Such information is often added implicitly, for example through the choice of derivative filters in neural imaging \cite{kato2015wholebrain, Chen2019}. Both the resulting state variables and the geometry and topology of their trajectories offer important, coordinate-invariant understanding of the processes generating the dynamics.  

 We applied our approach to the posture time series of the locomotor behavior of the roundworm {\em C. elegans} and found that the state space is spanned by a 7D basis of interpretable modes $\Gamma$ (Fig.~\ref{fig:fig3}), and their coefficients $X$, which are qualitatively similar for all worms in our foraging dataset. The basis is divided into three groups closely corresponding to forward, backward and turning locomotion. Biologically, these behaviors are linked to three classes of motor neurons: A and B ventral cord neurons which drive and backward and forward locomotion respectively, and sublateral motor neurons such as SMB and SMD which control deep body bends \cite{Cook2019}. Furthermore, excitatory classes of ventral cord motor neurons were recently reported to be capable of spontaneous rhythm generation and proposed to be central pattern generators for forward and backward locomotion \cite{Gao2018,Xu2018a}. We expect worms defective in different motor neurons to display a smaller projection on the behavioral modes in an interpretable manner. Although we have described the results for a 7D dimensional embedding, similar results are also found in 6 and 8 dimensions (Fig.~\ref{Fig:supp_w7_modes}-\ref{Fig:supp_ensembl_modes},\ref{Fig:supp_ensembl_m68}). Indeed, it is likely that higher modes capturing head movement and other subtle motions exist but carry little predictive structure in our analyzed conditions. Mutant worms defective in motor control, different sensorimotor contexts, or even faster sampling rates could reveal the presence of subtle, additional dynamics.

In our embedding, the state space trajectories retained significant variability, occupying much of the volume in the reconstructed space. A measure of this volume is the Kaplan-York dimension and we find $D_{KY}\sim 6$, not substantially smaller than the embedding dimension. One hypothesis for this variability is that worm behavior is stochastic and results from noise induced transitions between a finite number of elements such as stable limit cycles representing individual stereotyped motifs \cite{Revzen2011,stephens2011pnas,Berman:2014ba,Berman2016}. However, the exponential divergence of nearby state-space trajectories (Fig.~\ref{Fig:supp_divplot}) and the consistency of this divergence with the spectrum of unstable periodic orbits (Fig.~\ref{fig:fig4}), as well as the symmetric Lyapunov spectrum (Fig.~\ref{fig:fig5}) provide evidence for important, deterministic variation. From the perspective of deterministic chaos, behavioral dynamics are an aperiodic wandering among an infinite number of unstable periodic orbits, allowing an animal to generate an infinite number of behavioral sequences. Indeed, this agrees with the finding that the number of novel sequences in \emph{C. elegans} behavior grows with the observation time \cite{Schwarz2015}. On the other hand, stereotyped trajectories can emerge naturally as orbits with low values of the maximal Floquet exponent. Such trajectories can also be generated by stabilizing periodic orbits with control, e.g. a simple linear controller of the form $K(\textbf{g}(t)-\textbf{x}(t))$, where $\textbf{g}(t)$ are the desired goal dynamics, $\textbf{x}(t)$ is the current state and $K$ is a control gain matrix \cite{Pyragas1992, Jackson1997}.

The symmetric form of the Lyapunov spectrum suggests that the worm's behavioral dynamics can be interpreted as normal modes of a system of coupled, damped and driven, Hamiltonian oscillators,

\begin{equation}
\begin{aligned}
\dot{Q}_{i} & =\frac{\partial H}{\partial P_{i}}\\
\dot{P}_{i} & =-\frac{\partial H}{\partial Q_{i}} + C(Q_{i}, P_{i} ,\psi(t)),
\end{aligned}
\label{eqn:eqn-hamil}
\end{equation}
where $(Q_{i},P_{i})$ are the generalized position and momentum coordinates for the $i^{th}$ normal mode. The Hamiltonian is a scalar function governing the time-independent dynamics resulting from the mechanics of the worm's body, while $C(Q_{i}, P_{i} ,\psi(t))$ encapsulates the time-dependent neuromuscular control forces due to interaction of worm's body with the environment, proprioceptive feedback and neural processing of various sensory stimuli $\psi(t)$. Dynamics based on Hamiltonian structure are often associated with optimality and conservation laws and multiple efforts have reported quantities that remain roughly constant across a range external loads during \textit{C. elegans} locomotion, such as the normalized wave length of the body wave, angle of attack, bending power, and the phase relationship between the muscle activity and body curvature \cite{FangYen2010, Karbowski2006, Backholm2015, Berri2009, Butler2015}. Following the example of thermostatted dynamics (designed to capture constant temperature dynamics, see e.g.~\cite{Evans1986,Bright2005}), such emergent constants could be explained through feedback control arising from proprioceptive feedback, which is thought to underlie gait modulation in \textit{C. elegans} \cite{Boyle2012, Wen2012}. Our work also allows for connections between non-equilibrium thermodynamics and worm behavior. For example, worm dynamics breaks the Hamiltonian time-reversible symmetry in a continuous fashion via the dissipation rate $\alpha$, which sets the characteristic time-scale at which dynamics can be considered time-reversible symmetric. In addition, the sum of Lyapunov exponents reported here is an estimate of the entropy production rate \cite{Daems1999}. 

The dynamical invariants such as Lyapunov exponents, dimensions and entropies made accessible by our embedding approach provide important constraints and new understanding for short-time behavior consisting of neuromuscular control along with the biomechanics of the body and its environmental interaction. However, longer timescales are also present in the short periodic orbits, which are interpretable as forward/backward locomotion, and also longer time reorientation sequences such as pirouettes. Longer timescales can also be addressed through a systematic coarse-graining of the continuous state space dynamics which results in a transfer operator, see e.g.~\cite{Bollt2013}. In this approach the partition itself subsumes much of the nonlinearity so that the eigenvalues of the transfer operator can provide a systematic and useful timescale separation. In contrast, linear measures like the power spectrum are often not informative on the original dynamics of complex systems \cite{2019arXiv190811405R}. 

While we expect a dynamical systems perspective to be generally useful in understanding natural behavior, the analysis here benefits from the relative simplicity of the worm's foraging dynamics and the resulting interpretability of the modes.  Though other settings and organisms may generate more complex embeddings, important dynamical information such as trajectory stability and dynamical invariants can still be extracted from the state space reconstruction. Embedding ideas have also been recently used to understand the global brain dynamics of {\em C. elegans} \cite{Brennan2019} and to identify metastable sets and slow order parameters from molecular dynamics simulations using Markov operators \cite{Perez-Hernandez2013,Schwantes2013}.

Across wide areas of science there has been a remarkable increase in the availability of precise, multidimensional and dynamical data and new analysis ideas are emerging (see e.g.~\cite{Costa2019, Linderman2019, Daniels2019, BruntonKutz2019}). Here, we improve on the prior work on state space reconstruction \cite{Takens1981,Casdagli1991a,Stark2003,Muldoon1998,Stark1999,Huke2007,Deyle2011}, where much was in the context of either univariate measurements or known dynamical systems and included a heuristic search of reconstruction parameters. However, challenges associated with high-dimensionality, data sampling and nonstationarity remain. For example, the one-step error for $N$ samples from a $D$ dimensional dynamical system is $E(1) /e_{s} \approx N^{-1/D}$ \cite{Farmer1982,Farmer1987}-larger dimensional systems require exponentially more data to keep $E(1)/e_{s}\ll1$.  A related difficulty is the Euclidean metric used to find nearest neighbor distances, which is invalid even in lower-dimensional spaces with large curvature fluctuations. In this setting, it might be possible to use metric learning algorithms \cite{Xing2003} to recover a suitable metric from data.  Finally, recent progress in leveraging artificial neural networks to recover dynamical invariants \cite{Pathak2017} and to seek state-space representations with parsimonious dynamics \cite{Champion2019} offers promising directions for combining a principled dynamical perspective with high-dimensional, real-world systems.

\acknowledgments{We thank David Jordan, Ian Ehteridge and Antonio Celani for comments. This work was supported by OIST Graduate University (TA, GJS), a program grant from the Netherlands Organization for Scientific Research (AC, GJS), and by Vrije Universiteit Amsterdam (GJS).}

\section*{Methods}
{\noindent {\bf Software}:} Code for all analysis reported here was written in MATLAB \cite{MATLAB2017} and is publicly available: \url{https://bitbucket.org/tosifahamed/behavioral-state-space}. 

\medskip

{\noindent {\bf Experimental Details:}}  A brief description of the foraging and escape response datasets is given below. For more details  please see the original manuscripts \cite{Broekmans2016,stephens2011pnas}.

\medskip

{\noindent {\bf Foraging Dataset:}} $N=12$ L4-stage N2 worms were recorded at $32$\,Hz with high resolution tracking microscopy. For the analysis here the data was downsampled to 16\,Hz. Worms were cultivated under standard conditions at $20^{\circ} {\rm \, C}$ \cite{sulston1974}. Before the assay, worms were cleaned of \textit{E.coli} bacteria by a 1-minute immersion in NGM buffer. Worms were then placed on a $9.1$cm assay plate (Petri-Dish) with a $5$cm radius copper ring pressed into the agar surface for confinement. The assay started $5$ minutes after the transfer and lasted 35 minutes. 

\medskip

{\noindent {\bf Escape Response Dataset:}} $N=92$ mid to late L4 stage N2 worms were targeted on the head with a $100\, {\rm ms}$, $75 \, {\rm mA}$ IR laser pulse from a diode laser ($\lambda=1440$nm), resulting in a localized temperature change of approximately $0.5^{\circ} {\rm \, C}$. Images were recorded at $20 \, {\rm Hz}$ for $30 \, {\rm s}$ ($10 \, {\rm s}$ before stimulation and $20 \, {\rm s}$ after stimulation). To prevent adaptation each worm was only assayed once. To match the sampling rate of the foraging dataset, the posture time series was interpolated and downsampled to $16\, \rm {Hz}$ using the  MATLAB \cite{MATLAB2017} \texttt{resample} command.

\medskip

{\noindent {\bf State Space Reconstruction for the Lorenz system:}} 
We simulated the Lorenz system \cite{Lorenz1963},
\begin{equation*}
\begin{aligned}
\dot{s_{1}} & =10( s_{2} -s_{1})\\
\dot{s_{2}} & =s_1(28-s_3)-s_2\\
\dot{s_{3}} & =s_{1} s_{2} -\frac{8}{3} s_{3}
\end{aligned}
\label{eqn:lorenz}
\end{equation*}
using MATLAB's \texttt{ode45} Runge-Kutta ODE solver \cite{MATLAB2017} with a time-step $dt=0.01$\,s and error tolerances of $10^{-8}$. We take the variable $s_1$ as the observation time series $y(t)$. To simulate a noisy observation process we add to $y(t)$ a uniform white noise with standard deviation of $0.5\%$ the standard deviation of $s_1$. 

\

\medskip

{\noindent {\bf Image Analysis and Posture Space Estimation:}}
The tracking and posture space estimation follows \cite{Broekmans2016}. Briefly, we parameterize the shape of a worm by tangent angles calculated at 100 points along the body image skeleton. For a recording session of $T$ frames, this results in a ${T \times 100}$ matrix $\mathbf{\Theta}$, containing the shape information for each uncrossed frame where the worm's body does not intersect itself. Next, a $5$-dimensional approximation of the 100 dimensional posture space is calculated by projecting the elements of $\mathbf{\Theta}$ onto the basis given by the first $5$ singular vectors (eigenworms) of $\mathbf{\Theta}$. For frames with a body crossing an inverse tracking algorithm is used to identify the eigenworm projections \cite{Broekmans2016}.

\medskip

{\noindent {\bf Worm State Space Reconstruction:}}
Given a $d$-dimensional time-series in $Y=[\mathbf{y}_{1}^{1:T},\cdots,\mathbf{y}_{d}^{1:T}]$, along with an estimate of the optimal embedding window $K^{*}$, and minimum embedding dimension $m^{*}$, the state space reconstruction proceeds as follows. First, we create the $L \times K^{*}d$ matrix $\overline{Y}_{K^{*}}$ containing delayed copies of the mean subtracted measurements, $\overline{Y}_{K^{*}}=[\mathbf{y}_{1:d}^{1:L}, \mathbf{y}_{1:d}^{2:(L+1)}, \cdots, \mathbf{y}_{1:d}^{K:T}]$, where $L=(T-K^{*}+1)$. For the postures of \textit{C. elegans}, the measurements are composed of $d=5$ eigenworm coefficients. Next, we perform ICA on the space formed by the first $m^{*}$ singular vectors of $\overline{Y}_{K^{*}}$ using the FastICA algorithm \cite{Hyvaerinen1999} to obtain an $m^{*}$-dimensional state space spanned by the independent basis vectors, which we call behavioral modes and denote $\mathbf{\Gamma}$. Projections of $\overline{Y}_{K^{*}}$ on the state space are contained in the $L \times m^{*}$ state space matrix $X$. Each row, $\vec{x}(t)$, of $X$ is the behavioral state encoding the instantaneous behavior of the worm at time $t$, while the temporal sequence, $[\vec{x}(t),\cdots,\vec{x}(t+\tau)]$, forms a continuous trajectory in state space which encodes the shape change dynamics of a behavioral sequence. 

\medskip 

{\noindent {\bf Choosing Reconstruction Parameters by Maximizing Predictability:}} To choose the reconstruction parameters $(K,m)$ we first vary $K$ in the range $1 \leq K \leq K_{max}$ and estimate $T_{pred}$ in the candidate state space $\overline{Y}_{K}$ formed by the delayed observations. We set $K^{*}$ as the minimum $K$ where $T_{pred}$ as a function of $K$ begins to decrease. In cases where $T_{pred}$ saturates but doesn't decrease, we choose $K^{*}$ as the $K$ at which $T_{pred}$ saturates. As a guide, we choose $K_{max}$ such that for any delay larger than $K_{max}$ the autocorrelation function of the observations is close to zero. If $K_{max}$ appears too short, then it can be increased step-wise until $T_{pred}(K)$ starts decreasing. For the Lorenz system we have $K_{max}=100\,{\rm frames}$, while for the worm data we have $K_{max}=30\,{\rm frames}$. Intuitively, $K^{*}$ should allow the reconstruction to capture the fastest time scale of the system, which for chaotic systems is set by the period of the smallest UPO, $p_{min}$. Increasing $K^*$ further filters across longer periods and in the limit $K \to \infty$, the SVD filter becomes a discrete Fourier transform \cite{Vautard1989}. On the other end, $K^*$ should be large enough to embed the dynamics completely. Using the bound given by Takens embedding theorem \cite{Takens1981,broomhead86}, we get $(2m^{*}+1)/d \leq K^{*} < p_{min}$. Empirically, we find that $p_{min}/4 \leq K^{*}\leq p_{min}/2$.  

Once the embedding window is set as $K^{*}$, we next perform the singular value decomposition $\overline{Y}_{K^{*}}=U \Sigma V^{T}$. The first $m$ columns of $U$ contain the normalized projections of $\overline{Y}_{K^{*}}$ onto its first $m$ singular vectors. To find the embedding dimension, we vary $m$ and compute $T_{pred}$ as above. We set the embedding dimension $m^{*}$ as the minimum $m$ where $T_{pred}$ as a function of $m$ saturates or begins to decrease. 

\medskip

{\noindent {\bf Nearest Neighbor Prediction}:} We estimate the $\tau$-step future of an observation $\vec{y}(t^{\prime})$, denoted ${\vec{y}}_{\rm est}( t^{\prime}+\tau )$, from an average of the $\tau$-step future of $N_b$ nearest neighbors of the corresponding state space point $\vec{x} (t^{\prime})$. Specifically, we find $N_{b}$ nearest neighbors of $\vec{x}(t^{\prime})$ in state space, denoted by $\vec{z}(t^{\prime};r)$ for the $r^{th}$ nearest neighbor of $\vec {x}(t^{\prime})$, and average their values after $\tau$ steps, $\vec{x}_{\rm est}( t^{\prime}+\tau ) =\langle \vec{z}( t^{\prime}+\tau ;r) \rangle _{r}$, for all $N_{b}$ neighbors. Finally, we project $\vec{x}_{\rm est}(t^\prime+\tau)$ back to the observation space to get $\vec{y}_{\rm est}( t^{\prime}+\tau )$. 
We take only the transverse nearest neighbors (i.e. neighbors that are not in succession). The transverse nearest neighbors of $\vec{x}(t^{\prime})$ are identified by the local minima of $R_{t^{\prime}}(t)$ estimated using the \texttt{findpeaks} function in MATLAB \cite{MATLAB2017}, where $R_{t^{\prime}}(t)$ is the distance between $\vec{x}(t^{\prime})$ and all other points in state space. We quantify the $\tau$-step prediction accuracy by the root mean squared error
\begin{equation*}
E( \tau ) =\sum ^{d}_{j=1} \langle \left( y_{j}\left( t^{\prime} +\tau \right) -(y_{est})_{j}\left( t^{\prime} +\tau \right)\right)^{2} \rangle ^{1/2}_{t^{\prime}}
\end{equation*}
for $N=10^4$ different test points $\vec{y}(t^\prime)$ in the measurement time series. The predictions are made to a maximum prediction time which is long enough so that $E(\tau)$ saturates to $e_s$.  The root mean squared error is also a function of the $N_b$, the total number of nearest neighbors used for prediction. Making this dependence explicit, we write $E(\tau,N_{b})$ when $N_b$ is considered a variable. We set the number of neighbors by minimizing the bias and variance of the one-step prediction error.

\medskip

{\noindent {\bf Prediction Timescale $T_{pred}$}:} In cases where the error growth $E(\tau)$ is well approximated by a sigmoid (conjectured by Lorenz for chaotic systems with a single Lyapunov exponent $\lambda$ \cite{Lorenz1995}), one can show $T_{pred}=\displaystyle{\frac{1}{\lambda }\log\left(\frac{e_{s}}{e_{1}}\right)}$, where $e_1$ is the one time step error $E(1)$. Based on information theoretic considerations Farmer derived the upper bound for the predictability time scale as $\displaystyle{\frac{D_{I}}{h_{KS}}\log\left(\frac{e_{s}}{e_{1}}\right)}$, where $D_{I}$ is the information dimension and $h_{KS}$ is the Kolmogorov-Sinai entropy rate, which is consistent with our calculation for the sigmoid assumption. Importantly, these estimates shed light on the asymptotic behavior of $T_{pred}$. For small values of $K$ and $m$, the error is affected by some fraction of false nearest neighbors due to underembedding \cite{Kennel1992} leading to an overestimate of the local expansion rate and consequently the positive Lyapunov exponents. This causes a drop in $T_{pred}$ via the $1/\lambda$ term. On the other hand, as we increase $K$, the average Euclidean distance between nearest neighbors, $e_1$ steadily increases leading to a decrease in $T_{pred}$ for high dimensions. In the middle of these two extremes we find a range of suitable values for the embedding window $K$. The SVD coordinates are weighted by decreasing singular values, which correspond to the variance of the data projected along the different singular vectors. In the noiseless case, the singular values decay towards zero, while in the presence of noise they decay before saturating to the standard deviation of noise (termed noise floor in Ref. \cite{broomhead86}), thus higher dimensions are generically dominated by noise in this case. Consequently, $e_1$ doesn't increase as a function of $m$ in the noiseless case leading $T_{pred}$ to saturate after successful embedding. However, in the presence of noise $e_1$ increases causing $T_{pred}$ to go down.

\medskip

{\noindent {\bf Calculation of $T_{pred}$}:} We developed a fixed-point algorithm to estimate $T_{pred}$. We begin with an initial guess of $e_{s}$ labeled $e^{0}_{s}$ and time $\tau ^{0}_{s}$ such that $E( \tau )  >e^{0}_{s}$ for all $\tau \geq \tau ^{0}_{s}$. Next, noting that for large times $\int E(\tau )\,d\tau = e_{s}\tau-\Delta$ we fit a line to to a numerical estimate of $\int E(\tau )\,d\tau$ from $\tau^{0}_{s}$ to $\tau_{max}$ . The slope of this line provides the next estimate of $e_{s}$ labeled $e^{1}_{s}$, and the intercept is the next estimate of the area $\Delta$, labeled $\Delta^{1}$. We use $e^{1}_{s}$ to again estimate $\tau ^{1}_{s}$ and fit a line to $\int E(\tau )\,d\tau$ from $\tau ^{1}_{s}$ to $\tau_{max}$, repeating the process until the estimates for $e^{j}_{s}$ and $\Delta^{j}$ converge. Using the final estimates of $\Delta$ and $e_s$ we can get a robust estimate $\displaystyle T_{pred}=\frac{\Delta}{e_s}$. A schematic of this iterative process is shown in Fig.~\ref{Fig:S6}. In our experience it only takes 3-4 iterations for the estimates to converge. To obtain the error bars we bootstrap across the prediction test points, generating $100$ bootstrapped $E(\tau)$ curves along with $T_{pred}$ estimates for each. These are then used to estimate the $95\%$ confidence intervals of $T_{pred}$. 

\medskip

{\noindent {\bf Periodic Orbits}:} To detect periodic orbits of length $p$ we identify close recurrences in the state space after approximately $p$ time steps \cite{Lathrop1989,Pawelzik1991,Badii1994}. Specifically, we start at point $\vec{x}(i)$ in state space, and find smallest $k >i$ such that $\| \vec{x}( k) -\vec{x}(i) \| < \epsilon$. The sequence $[\vec{x}( i) ,\vec{x}( i+1) ,\cdots ,\vec{x}(k)]$ is then stored as a periodic orbit of period $p=k-i$. Again, we only consider transverse recurrences to avoid sequential points. If such a $k$ cannot be found then it implies that no periodic orbits exist at the scale of $\epsilon$. The distance scale of the recurrence $\epsilon$ is calculated through the function $\epsilon(r,t)$ as defined in Ref. \cite{Pawelzik1991}, which gives the $r^{th}$ smallest distance between state space points separated by time $t$. An example of this function is shown in Fig.~\ref{Fig:supp_fwd_rev_upo}A. Small values of $\epsilon(r,t)$, identified by local minima of $\epsilon(r,t)$, indicate close recurrences after times $t_{*}$, and in consequence reveal the existence of periodic orbits of length $t_{*}$. If $\epsilon(r,t)$ does not show any local minima, then periodic orbits cannot be detected from the data. In this manner $\epsilon(r,t_{*})$ gives the minimum distance at which we must look to find a periodic orbit of length $t_{*}$. The smallest recurrence time corresponding to the first local minima of $\epsilon(r,t)$ equals the smallest period $p_{min}$ detected in the data. In the Jacobian and maximal exponent calculations described below, $\epsilon_{*}$ is the distance corresponding to $p_{min}$. Finally, we set $r=m^{*}$, the dimension of the reconstructed state space.

\medskip

{\noindent {\bf Maximal Lyapunov Exponent $\lambda_{max}$}:} Our test for the exponential divergence of neighboring trajectories follows standard approaches \cite{Kantz1994}. Specifically, we consider a reference trajectory $\vec x(t^{\prime}+\tau)$ and its nearest neighbors within a distance $\epsilon_{*}$ (see Methods: Periodic Orbits). We then track the average distance between the reference and neighboring trajectories over time to obtain the curve $\delta_{t^{\prime}}(\tau)$. A significant linear region in the $\langle \log \delta_{t^{\prime}}(\tau) \rangle_{t^{\prime}}$ curve indicates an exponential divergence of neighboring trajectories, while the slope of the linear region provides an estimate of the maximal Lyapunov exponent $\lambda_{max}$. There is typically a transient before the exponential growth where the perturbation vector aligns itself with the Lyapunov vector corresponding to the maximal exponent. In the Lorenz system it is seen that this transient arises from the finite-size of the perturbation and vanishes in the infinitesimal limit. To avoid effects due to non-stationarities we perform this calculation on the final two minutes of the recording.

\medskip

{\noindent {\bf Jacobian Estimation}:} The Jacobian at the point $\vec{x}(i)$ in state space, denoted $\mathbf{J}_{\vec{x}(i)}$, is the derivative of the dynamics at $\vec{x}(i)$, forming the local linear approximation of the dynamics at that point. We use a modified version of the Jacobian estimation algorithm described in Ref. \cite{Deyle2016}, which solves a weighted regression problem $\mathbf{B}^{1}_{\vec{x}(i)} =\mathbf{B}_{\vec{x}(i)} \cdot \mathbf{J}_{\vec{x}(i)}$, where points are assigned weights according to their distance from $\vec{x}(i)$ as per the weighting function defined below. $\mathbf{B}_{\textbf{x}(i)}$ is a $(T-K^{*})\times (m+1)$ matrix containing all weighted state space points concatenated with a column of ones, while $\mathbf{B}^{1}_{\vec{x}(i)}$ is a $(T-K^{*}) \times m$ matrix containing all weighted successors. Each row of $\mathbf{B}_{\vec{x}(i)}$ and $\mathbf{B}^{1}_{\vec{x}(i)}$ is weighted by $w(k)=\exp\{-\frac{\| \vec{x}(i)-\vec{x}(k)\|}{\epsilon_{*}}\}$. The estimated local Jacobian matrix is then given by $\mathbf{J}_{\vec{x}(i)} =\mathbf{B}^{\dagger}_{\vec{x}( t)} \cdot \mathbf{B}^{1}_{\vec{x}(i)}$ where $\mathbf{B}^{\dagger }_{\vec{x}(i)}$ is the pseudoinverse of $\mathbf{B}_{\vec {x}(t)}$ which we compute using the \texttt{pinv} function in MATLAB \cite{MATLAB2017}. Note that $\epsilon_{*}$ is the distance scale corresponding to the minimum period recurrence (see Methods: Periodic Orbits).

\medskip

{\noindent {\bf Floquet Exponents}:} The real parts of the Floquet exponents of a periodic orbit, which measure their stability, are equal to the Lyapunov exponents of the orbit \cite{Guckenheimer1983}. To estimate the Floquet exponents of a periodic orbit, we estimate the maximal local Lyapunov exponent along the orbit using established algorithms \cite{Abarbanel91variation,Abarbanel_1992}. We use a  recursive QR iteration for obtaining the eigenvalues of the product of Jacobian matrices along a periodic orbit. The average of the logarithms of the eigenvalues give the $m$ local Lyapunov exponents of the orbit in an $m$ dimensional state space. The maximal exponent is then the Floquet exponent of the periodic orbit.

\medskip

\begin{figure*}
\begin{center}
\includegraphics[width=0.9\textwidth]{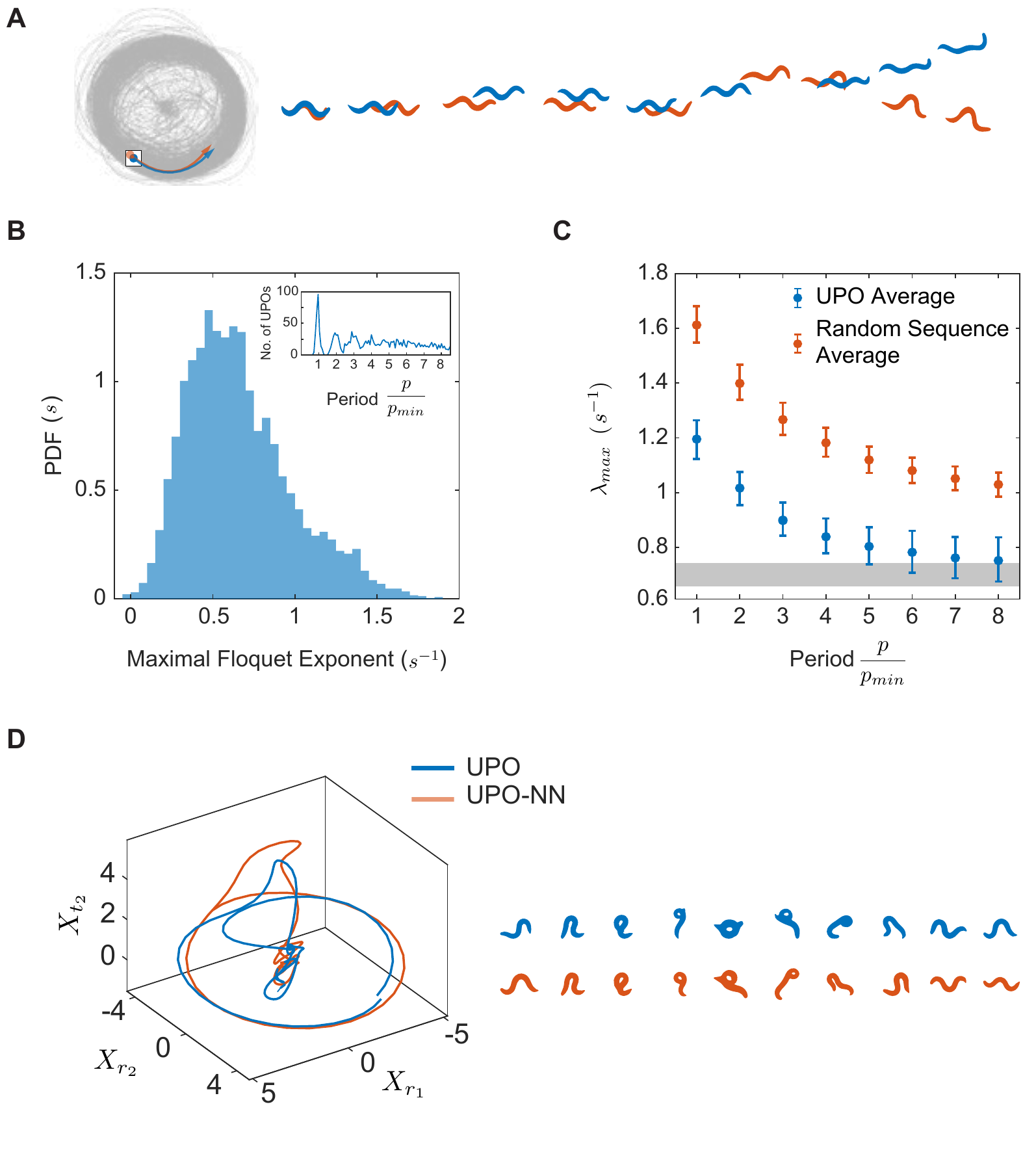}
\caption{{\bf Chaotic State Space Dynamics Reveal a Deterministic Component of Behavioral Variability} \textbf{(A)} Neighboring state space points correspond to similar posture sequences, but the futures of these points diverge at longer times. \textbf{(B)} We identify periodic orbits in the state space using the ensemble embedding and quantify the stability of each periodic orbit through its maximal Floquet exponent. The identified periodic orbits are predominantly unstable, as also observed in known chaotic systems. \textbf{(B, inset)} Histogram of the number of periodic orbits across all worms as a function of their period exhibits peaks at integer multiples of a minimum period $p_{min}$ corresponding to the period of a body wave during forward locomotion. \textbf{(C)}  We recover the characteristic divergence of the trajectories $\lambda_{max}$ (grey bar) by a weighted average of the maximal Floquet exponent of across orbits (blue). We quantify the state-space divergence across worms, Fig.~\ref{Fig:supp_divplot}, and find an exponential regime, yielding an average maximal Lyapunov exponent  $\lambda_{max}=0.69\,(0.65,0.74)\,s^{-1}$. In contrast, an average from random sequences of the same length converges more slowly (red). \textbf{(D)} We show an example period-4 orbit composed of a sequence of reversal, deep body bend, and forward movement (blue). This same orbit is revisited at a later time (orange), resulting in a similar sequence of postures for these two long trajectories (right). 
}

\label{fig:fig4}
\end{center}
\end{figure*}

\begin{figure*}
\begin{center}


\includegraphics[width=0.9\textwidth]{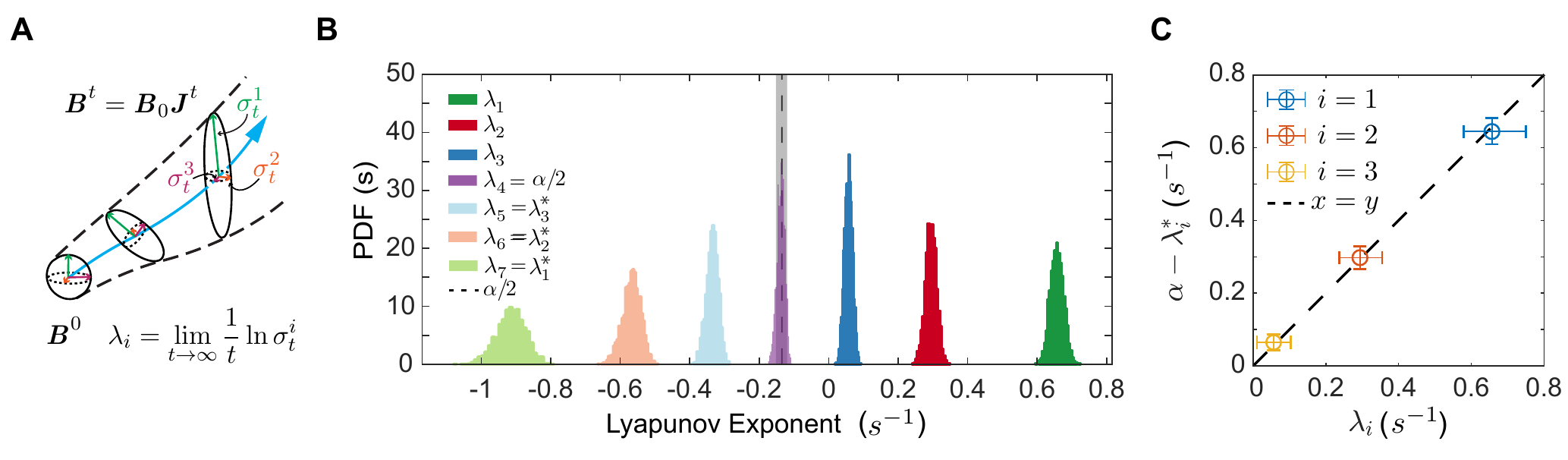}
\caption{{\bf Symmetric Lyapunov Spectrum Indicates Damped-Driven Hamiltonian Dynamics}  \textbf{(A)} The contraction and expansion of local volumes, suitably averaged across the full state space, provides important dynamical invariants and is quantified by a collection of exponents: the Lyapunov spectrum. We sketch the transformation of an initial volume $B^0$ to a new volume $B^t$ by the flow (blue curve) and the local Jacobian $J^t$. The spectrum of Lyapunov exponents is given by the size of the principal axes of $B^t$ for $t\rightarrow \infty$. \textbf{(B)} The Lyapunov spectrum for {\em C. elegans} foraging dynamics where the distribution for each exponent is constructed from bootstrap samples across different worms. We find two positive (chaotic) exponents driving variability, a near-zero exponent indicative of continuous (non-noisy) dynamics and four negative exponents which drive stereotypy. The spectrum is symmetric about the point $\alpha/2$ which coincides with $\lambda_{4}$. \textbf{(C)} Lyapunov exponents come as conjugate pairs that sum to $\alpha$, a symmetry suggestive of damped-driven Hamiltonian system.}
\label{fig:fig5}
\end{center}
\end{figure*}

{\noindent {\bf Lyapunov Exponents of Random Sequences}:} To calculate the exponents for short random sequences in Fig.~\ref{fig:fig4}C, we proceed as above but instead of the sequence being a periodic orbit, it is formed by starting at a random random point in state space and following it for the duration of a periodic orbit. As a result these sequences are not necessarily recurrent.

\medskip

{\noindent {\bf Lyapunov Spectrum}:} To estimate the full spectrum of Lyapunov exponents for a given state space, we use the recursive QR procedure as above \cite{Abarbanel91variation,Abarbanel_1992}, but now calculate the product of Jacobians over the entire recording instead of a short UPO segment.

\medskip 

\bibliographystyle{apsrev4-1_PRX_style}
\bibliography{Bibliography}


\clearpage


\onecolumngrid

\begin{center}
\textbf{\Large Supplementary Material}
\end{center}

\setcounter{figure}{0}
\setcounter{page}{1}
\makeatletter
\renewcommand{\theequation}{S\arabic{equation}}
\renewcommand{\thefigure}{S\arabic{figure}}

\addtolength{\topmargin}{0.15in}
\addtolength{\textheight}{-0.15in}

\begin{figure}[htp]
\begin{center}
\includegraphics[width=0.9\textwidth]{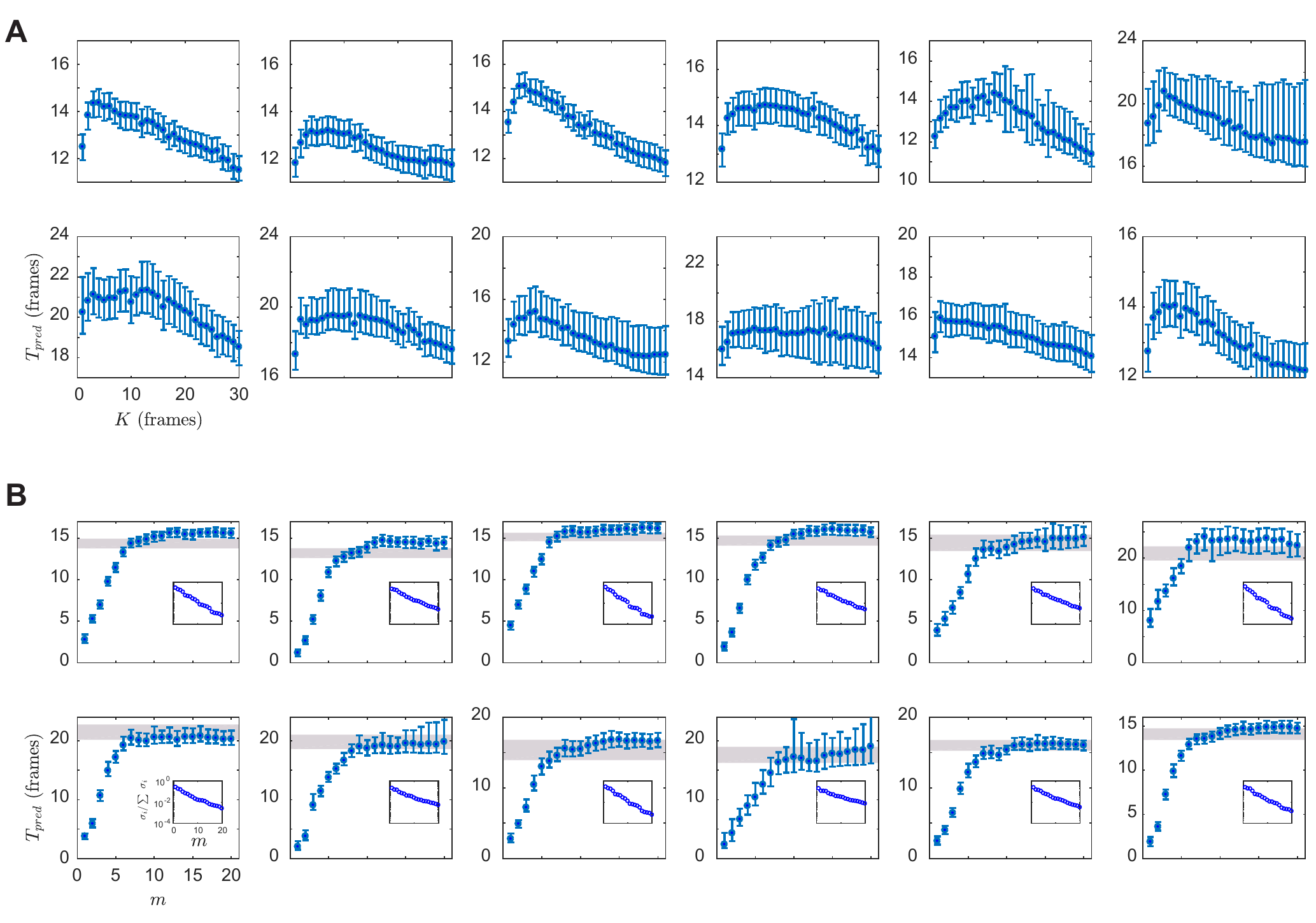}
\caption{{\bf State space embedding for all the worms in the foraging dataset.} Individual worm estimates of $T_{pred}$ as a function of $K$ are shown in \textbf{(A)}, and as a function of $m$ in \textbf{(B)}. We find similar curves across worms, despite the differences in their detailed dynamics.  Note that while the distance metric in the SVD space (B) and the space of delays (A) is different, which could result in inconsistencies in the estimation of $T_{pred}$, we find only minor differences between the maximum $T_{pred}$ in the two cases (gray bar in B). Inset shows the normalized singular value spectrum, which does not have a clear cut-off for any worm.
}
\label{Fig:sup_wwise_tpred}
\end{center}
\end{figure}

\begin{figure}[htp]
	\begin{center}
		\includegraphics[width=0.9\textwidth]{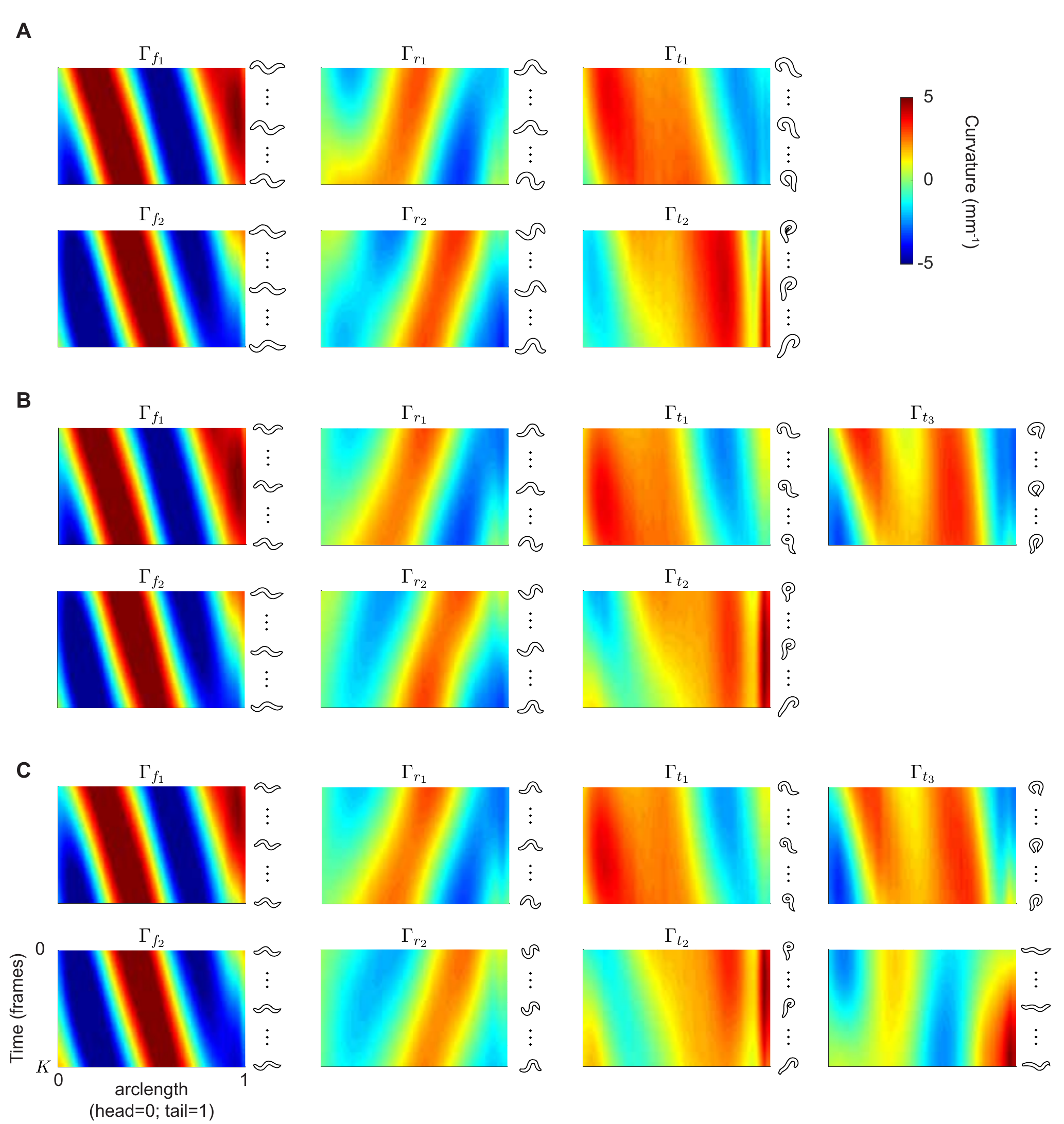}
		\caption{{\bf Dominant state space modes are stable across different embedding dimensions, with distinct groups independently capturing forward, reversal and turning behaviors} \textbf{(A-C)} Behavioral modes estimated for the worm in Fig.~\ref{fig:fig3} for dimensions, $m=6$ (A), $m=7$ (B), and $m=8$ (C); embedding window is set to $K=12$ frames. The modes retain their interpretability across dimensions. In a 6-dimensional embedding, there are two forward, two backward and two turning modes. In 7 dimensions one of the turning modes further splits into an omega-turn like mode ($\Gamma_{t_3}$) and a delta-turn like mode ($\Gamma_{t_1}$), while $\Gamma_{t_2}$ changes little. Furthermore, the reversal modes are more separable in 7 dimensions. The 8-dimensional state space retains the forward, reversal and turning dynamics along with an additional and subtle head-bending.
		}
		\label{Fig:supp_w7_modes}
	\end{center}
\end{figure}

\begin{figure}[htp]
	\begin{center}
		\includegraphics[width=0.85\textwidth]{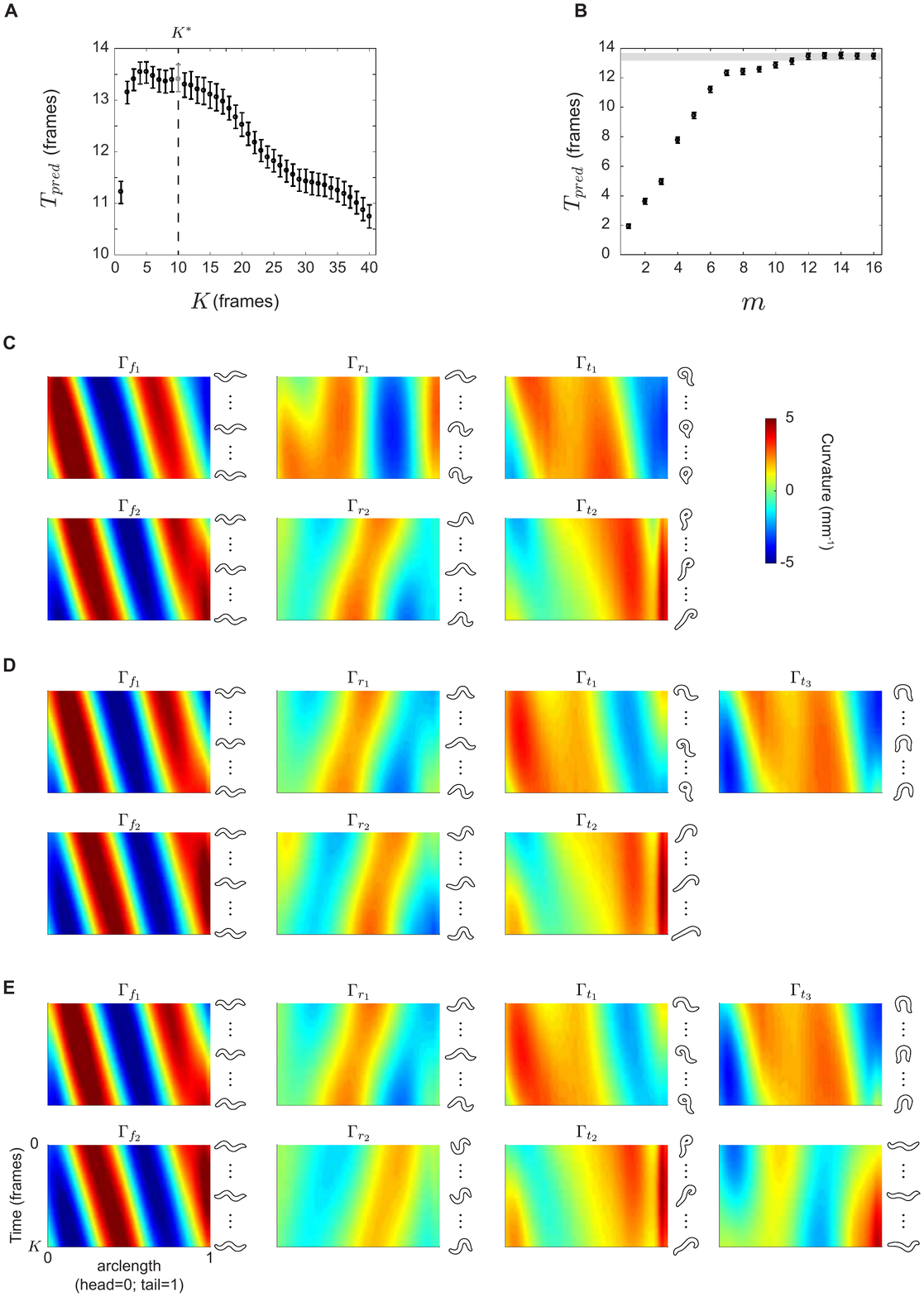}
		\caption{{\bf The ensemble embedding across all (N=12) worms is constructed from their concatenated posture time series and characterized by $K^{*}=10$ and $m^*=7$.} \textbf{(A-B)} $T_{pred}$ as a function of $K$ and $m$. We set $K^{*}=10$, approximately when $T_{pred}$ begins to decrease, and show  $T_{pred}(m)$ at this $K^*$. We show the resulting modes for $m=6$ (\textbf{C}), $m=7$ (\textbf{D}), and $m=8$ (\textbf{E}), and these are qualitatively similar to those obtained from our representative worm, Fig.~\ref{Fig:supp_w7_modes}. The additional modes present for embeddings greater than $m*=7$ offer only much smaller improvement in predictability and appear independent of forward, reversal and turning behaviors.
		}
		\label{Fig:supp_ensembl_modes}
	\end{center}
\end{figure}

\begin{figure}[htp]
	\begin{center}
		\includegraphics[width=0.9\textwidth]{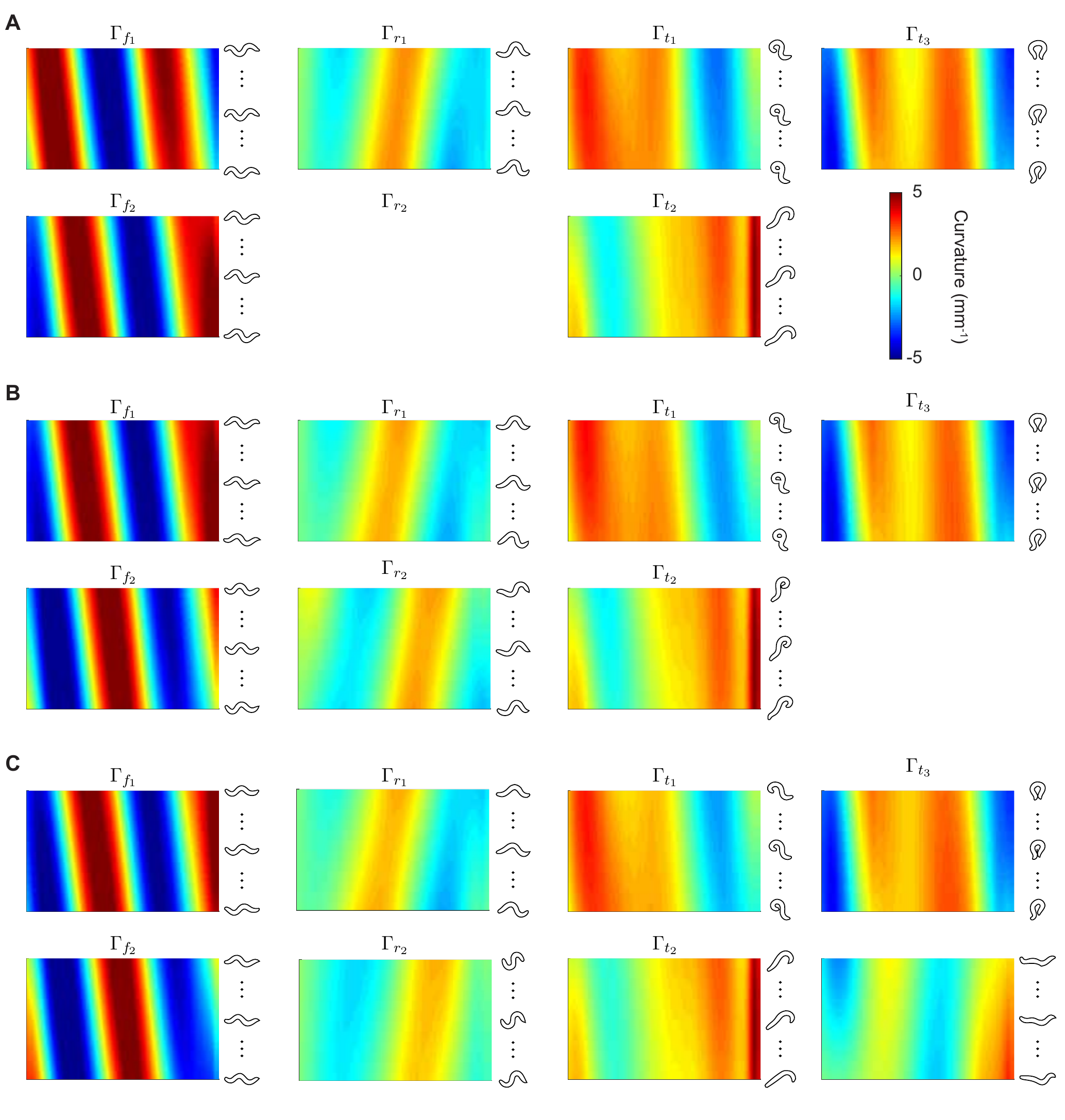}
		\caption{{\bf Ensemble embedding for different values of $K$ and $m$} \textbf{(A-C)} Behavioral modes estimated from the ensemble for $K=5$, and dimensions, $m=6$ (A), $m=7$ (B), and $m=8$ (C). \textbf{(D-F)} Same as above but for $K=15$. The modes are qualitatively similar across this variation.
		}
		\label{Fig:supp_ensembl_m68}
	\end{center}
\end{figure}

\begin{figure}[htp]
	\begin{center}
		\includegraphics[width=0.5\textwidth]{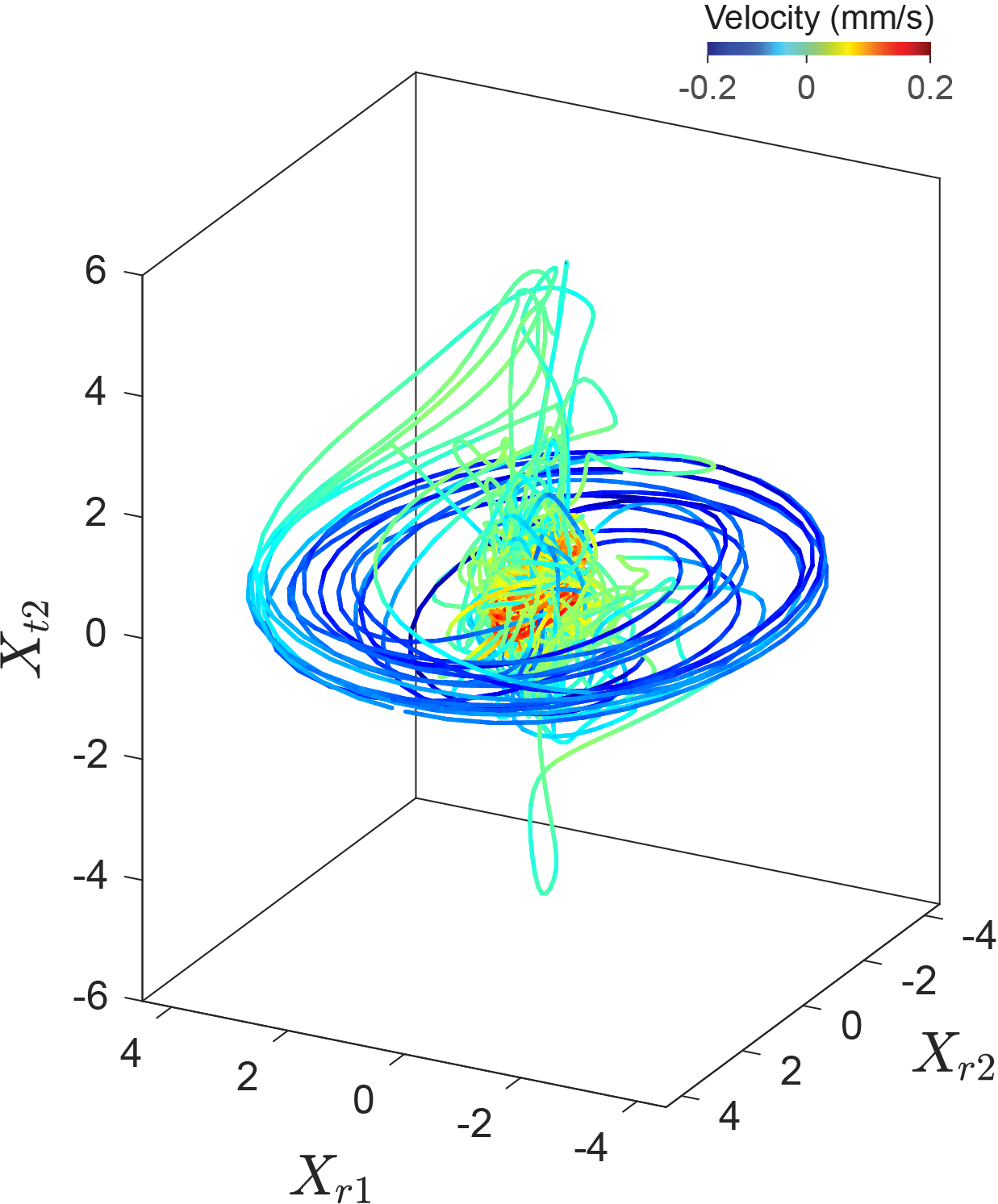}
		\caption{State space captures a commonly observed sequence where long reversals transition to forward via a deep body bends seen here as a large excitation in $X_{t2}$ as the reversal ends. Here we see that the blue (backward) and red (forward) bundles are smoothly connected via a large transient along the turning mode $X_{t_2}$
		}
		\label{Fig:supp_3dproj}
	\end{center}
\end{figure}

\begin{figure}[htp]
	\begin{center}
		\includegraphics[width=0.9\textwidth]{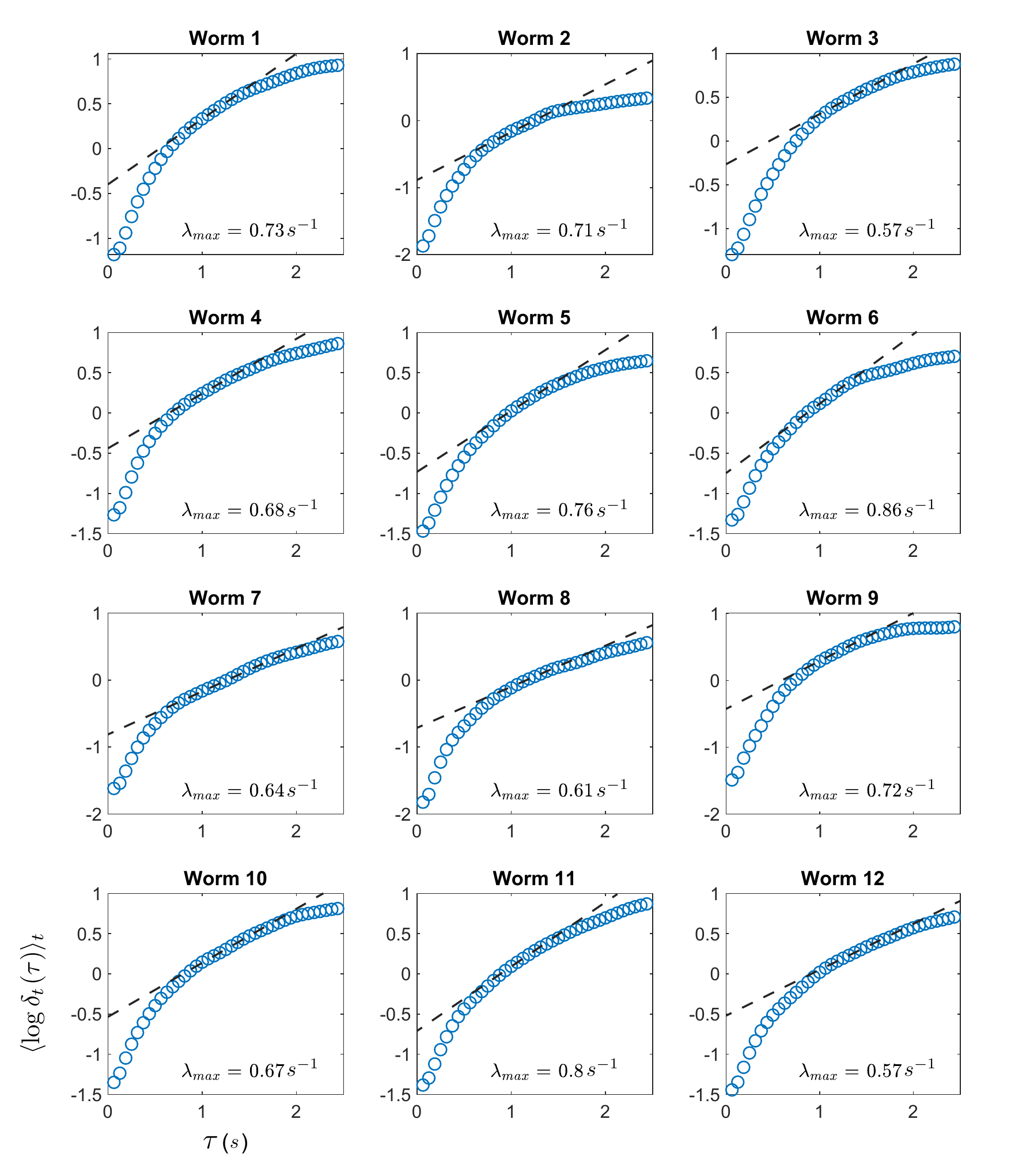}
		\caption{To quantify the state space divergence we plot the logarithm of the average distance between a trajectory and it's nearest neighbors, averaged over several starting reference trajectories. For each worm we find that after a transient, there is linear region showing exponential divergence.  The slope of the linear region provides an estimate of the maximal Lyapunov exponent $\lambda_{max}$ and the positive exponents are an indication of chaos in worm behavior.
		}
		\label{Fig:supp_divplot}
	\end{center}
\end{figure}

\begin{figure}[htp]
	\begin{center}
		\includegraphics[width=0.9\textwidth]{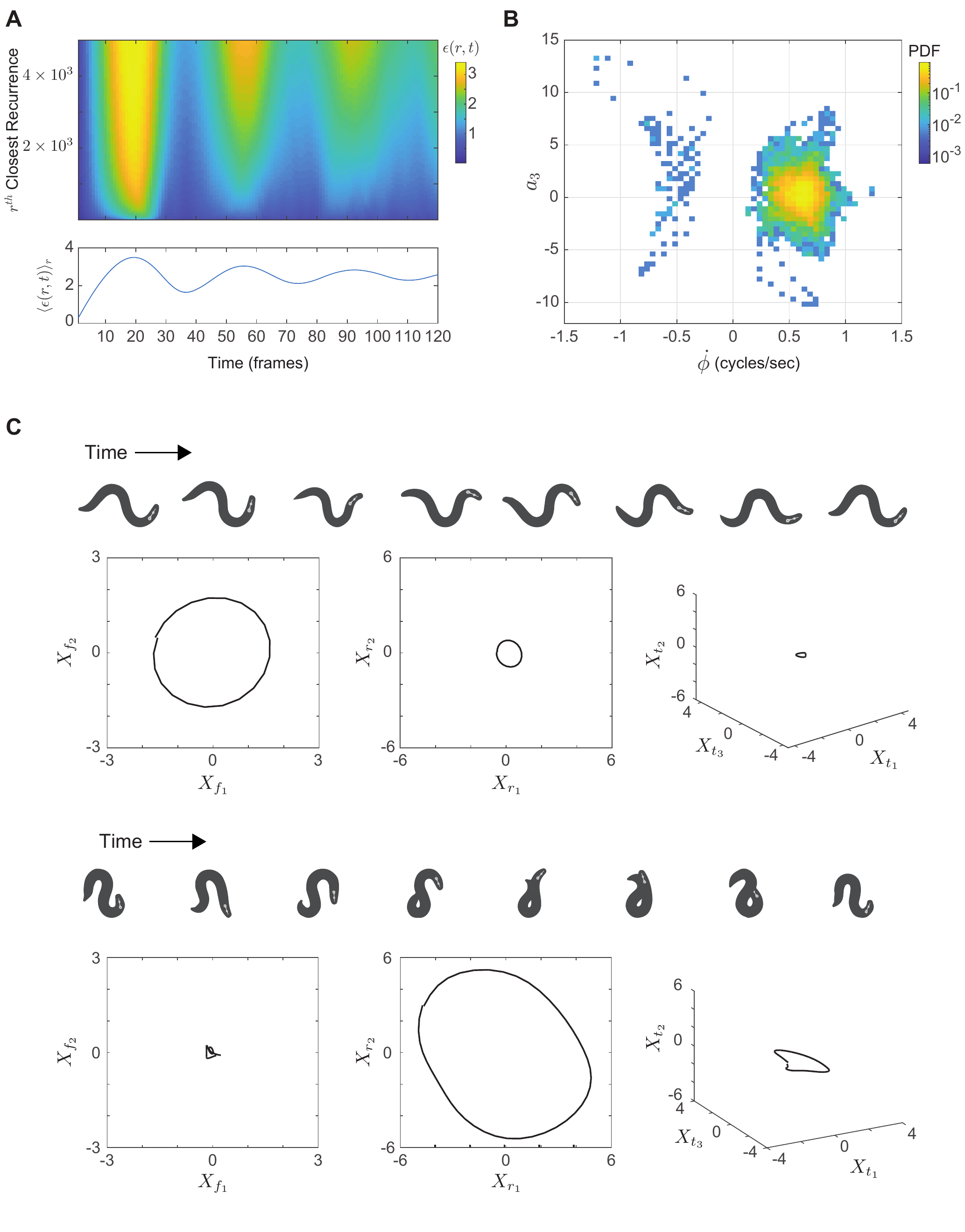}
		\caption{\textbf{(A)} The recurrence function $\epsilon(r,t)$ from the same worm in Fig~\ref{fig:fig3} for 120 frames and 5000 closest recurrences (top). Local minima of this function at times $t_{*}$, as seen in the average $\langle \epsilon(r,t) \rangle _r$  shown below correspond to close recurrences and identify periodic orbits of length $t_{*}$. The first local minimum is the smallest period $p_{min}$, which is $37$ frames in this example. For a given value of $r$, $\epsilon(r,t_{*})$ gives the distance threshold at which we must look to find a periodic orbit of length $t_{*}$. \textbf{(B)} Probability distribution of phase velocities $\dot{\phi}$ and third eigenworm coefficient $a_3$, which is proportional to mean body curvature, across all period-1 orbits of duration $p_{min}$ from all worms in the dataset. We see two clusters corresponding to forward and backward locomotion, as well as orbits with a dorsal or ventral bias (e.g.~orbits at bottom right and top left). \textbf{(C)} Example period-1 orbits from the same worm in (A) corresponding to forward (top) and backward (bottom) locomotion.
		}
		\label{Fig:supp_fwd_rev_upo}
	\end{center}
\end{figure}

\begin{figure}[htp]
	\begin{center}
		\includegraphics[width=0.9\textwidth]{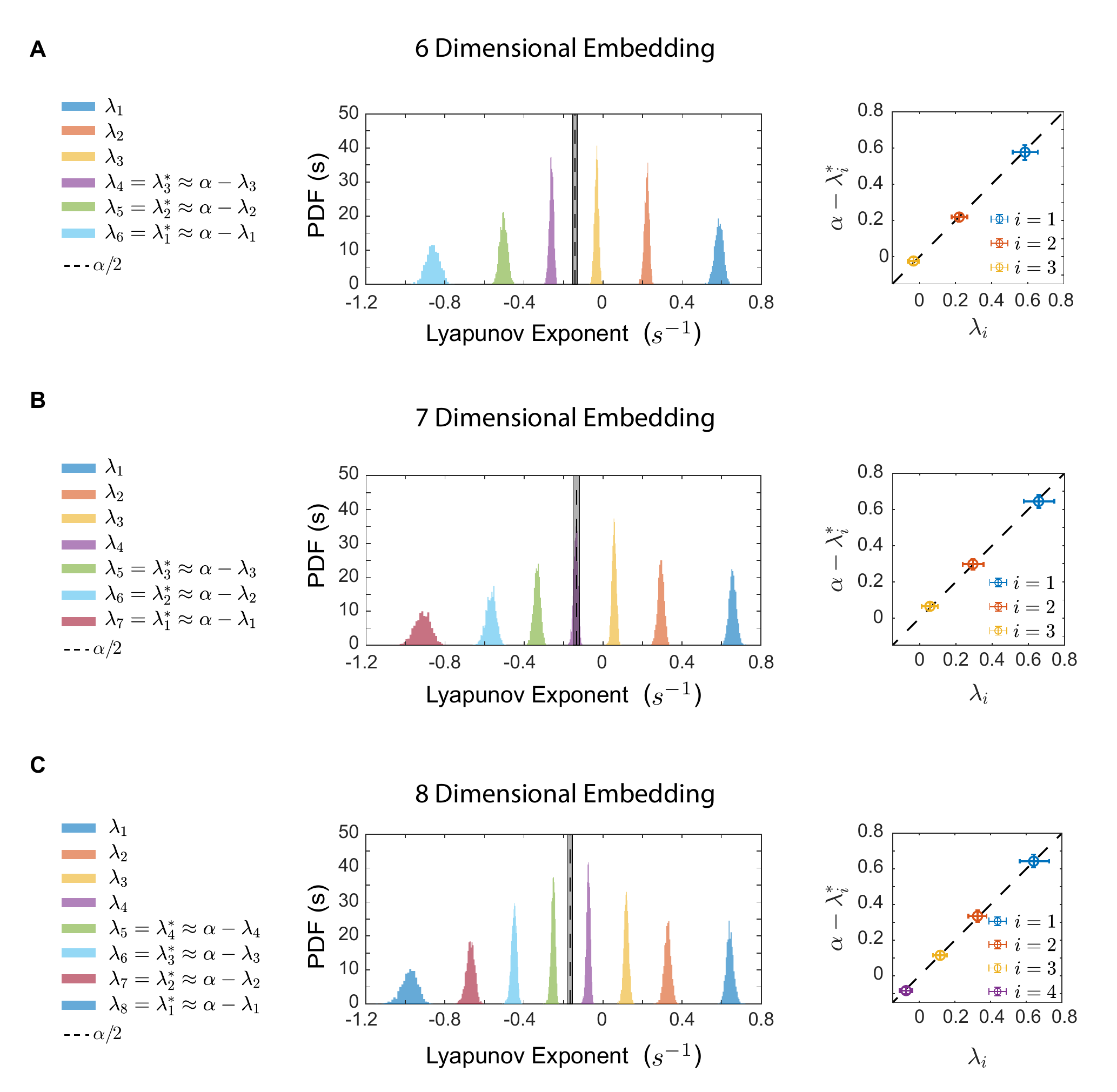}
		\caption{Lyapunov spectra computed from reconstructions of worm behavior in different embedding dimensions. Conjugate pairing of Lyapunov exponents is robustly observed in dimensions 6 and above.
		}
		\label{Fig:S5}
	\end{center}
\end{figure}

\begin{figure}[htp]
	\begin{center}
		\includegraphics[width=0.9\textwidth]{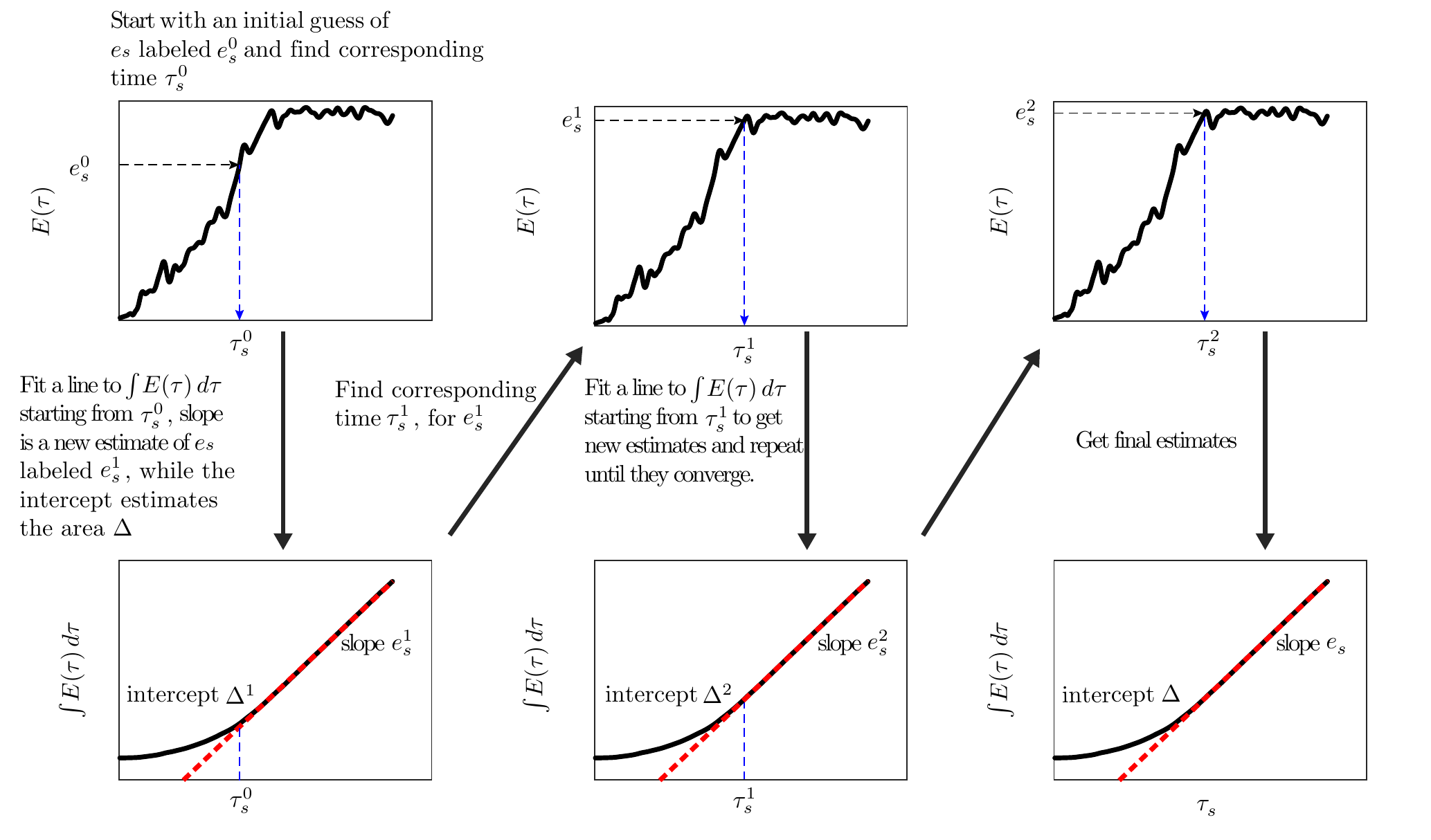}
		\caption{A schematic showing the fixed point algorithm for robust estimation of the asymptote $e_s$, and the area $\Delta$.
		}
		\label{Fig:S6}
	\end{center}
\end{figure}

\end{document}